\documentclass[aps,prl, twocolumn, superscriptaddress, longbibliography]{revtex4-2}

\usepackage{times}
\usepackage{siunitx}
\usepackage{braket}
\usepackage{amsmath}
\usepackage{amssymb}
\usepackage{graphicx}
\usepackage[caption=false, position=top, singlelinecheck=off, justification=raggedright]{subfig}
\usepackage{color}
\usepackage{pst-node}
\usepackage[unicode]{hyperref}
\hypersetup{unicode=true, colorlinks=true, linkcolor=blue, citecolor=blue}
\usepackage{float}
\usepackage[normalem]{ulem}

\begin{document}

\preprint{APS/123-QED}

\title{Observation of a zero-energy excitation mode in the open Dicke model}

\author{Anton B\"{o}lian}
\thanks{These authors have contributed equally to this work.}
\affiliation{Institute for Quantum Physics, Universit\"at Hamburg, 22761 Hamburg, Germany}

\author{Phatthamon Kongkhambut}
\thanks{These authors have contributed equally to this work.}
\affiliation{Institute for Quantum Physics, Universit\"at Hamburg, 22761 Hamburg, Germany}
\affiliation{Quantum Simulation Research Laboratory, Department of Physics and Materials Science, Faculty of Science, Chiang Mai University, Chiang Mai, 50200, Thailand}
\affiliation{Thailand Center of Excellence in Physics, Office of the Permanent Secretary, Ministry of Higher Education, Science, Research and Innovation, Thailand}

\author{Christoph Georges}
\affiliation{Institute for Quantum Physics, Universit\"at Hamburg, 22761 Hamburg, Germany}

\author{Roy D. Jara Jr.}
\affiliation{National Institute of Physics, University of the Philippines, Diliman, Quezon City 1101, Philippines}

\author{Jos\'{e} Vargas}
\affiliation{Institute for Quantum Physics, Universit\"at Hamburg, 22761 Hamburg, Germany}

\author{Jens Klinder}
\affiliation{Institute for Quantum Physics, Universit\"at Hamburg, 22761 Hamburg, Germany}

\author{Jayson G. Cosme}
\affiliation{National Institute of Physics, University of the Philippines, Diliman, Quezon City 1101, Philippines}

\author{Hans Ke{\ss}ler}
\email[]{hkessler@physnet.uni-hamburg.de}
\affiliation{Institute for Quantum Physics, Universit\"at Hamburg, 22761 Hamburg, Germany}

\author{Andreas Hemmerich}
\email[]{andreas.hemmerich@uni-hamburg.de}
\affiliation{Institute for Quantum Physics, Universit\"at Hamburg, 22761 Hamburg, Germany}

\date{\today}

\begin{abstract}
Approaching phase boundaries in many-body systems can give rise to intriguing signatures in their excitation spectra. Here, we explore the excitation spectrum of a Bose-Einstein condensate strongly coupled to an optical cavity and pumped by an optical standing wave, which simulates the famous Dicke-Hepp-Lieb phase transition of the open Dicke model with dissipation arising due to photon leakage from the cavity. For weak dissipation, the excitation spectrum displays two strongly polaritonic modes. Close to the phase boundary, we observe an intriguing regime where the lower-energetic of these modes, instead of showing the expected roton-type mode softening, is found to approach and persist at zero energy, well before the critical pump strength for the Dicke-Hepp-Lieb transition boundary is reached. Hence, a peculiar situation arises, where an excitation is possible at zero energy cost, but nevertheless no instability of the system is created. 

\end{abstract}

\maketitle
\begin{figure*}[ht]
\centering
\includegraphics[width=\linewidth]{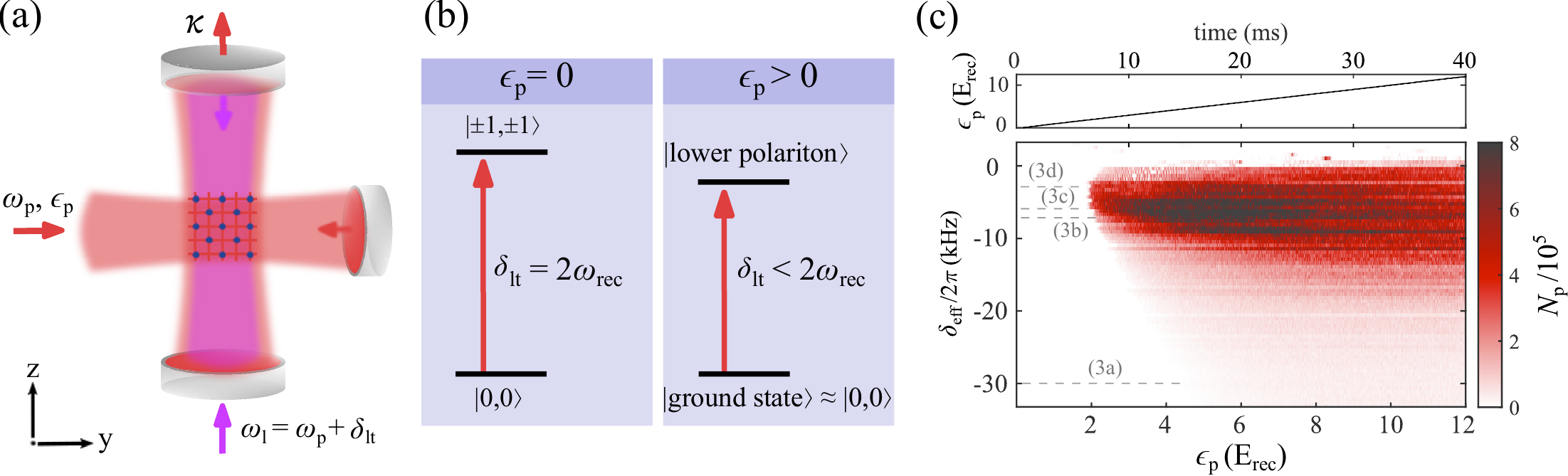}
\caption{(a) Illustration of the Bragg spectroscopy scheme used to probe the excitation energy of the polariton modes. Scans of the frequency detuning $\textit{$\delta$}_\mathrm{lt}$ between the on-axis longitudinal probe and the transverse pump waves are applied with variable fixed frequency $\textit{$\omega$}_\mathrm{p}$ and strength $\textit{$\epsilon$}_\mathrm{p}$ of the transverse pump. (b) Sketch of the mode softening of the lower polariton mode as the transverse pump strength increases. (c) Intracavity photon number $N_\mathrm{p}$ plotted against the effective detuning $\textit{$\delta$}_\mathrm{eff}$ between the transverse pump and the cavity and the transverse pump strength $\textit{$\epsilon$}_\mathrm{p}$. The detunings at which the excitation spectra shown in Fig.~\ref{fig:3} were measured ($\textit{$\delta$}_\mathrm{eff}=2\textit{$\pi$}\times\{-30,-7.5,-6,-3\}$ \si{\kilo Hz}) are marked accordingly. For negative effective detunings $\textit{$\delta$}_\mathrm{eff}<0$, the formation of a finite intracavity light field with photon number $N_\mathrm{p}$, arising above a critical value of the transverse pump strength $\textit{$\epsilon$}_\mathrm{p}$, indicates that the system self-organises into a superradiant phase.}
\label{fig:1}
\end{figure*}

The coupling of a quantum gas to an optical cavity provides an ideal experimental scenario to study the effects of strong light-matter interactions \cite{RevModPhys.85.553, Mivehvar_2021}, including cavity-induced cooling \cite{PhysRevLett.79.4974, PhysRevLett.84.3787, Maunz_2004}, cavity optomechanics \cite{PhysRevLett.105.133602, RevModPhys.86.1391, Brennecke_2008, PhysRevA.99.013849, Keszler_2014}, and quantum simulations of many-body physics \cite{PhysRevLett.125.060402, PhysRevX.8.011002, Dissipativetimecrystal, Continuoustimecrystal, Klinder_2015, Landig_2016}. Atom-cavity systems can also be used to emulate the open Dicke model \cite{PhysRev.93.99}, in which an ensemble of identical two-level systems is coupled to a single mode of the electromagnetic field. In cavity QED, the open Dicke model is typically realised by coupling the atomic ensemble to a single cavity mode, while the system interacts with the environment via photons leaking out of the cavity with the cavity field decay rate $\textit{$\kappa$}$. This has been achieved with thermal atoms \cite{PhysRevLett.91.203001} and both fermionic \cite{Piazza_2014, PhysRevLett.112.143002, PhysRevLett.106.210401, Helson_2023} and bosonic \cite{Baumann_2010, PhysRevLett.104.130401, doi:10.1073/pnas.1417132112} quantum gases. In the recoil-resolved regime, when the cavity bandwidth is smaller than the recoil frequency $\omega_\mathrm{rec}$, atom-cavity systems provide access to several distinctive phenomena, such as discrete \cite{Gong_2018, Dissipativetimecrystal, PhysRevA.92.023815, Kongkhambut_2024} and continuous \cite{Continuoustimecrystal, Skulte_2024, PhysRevA.99.053605, PhysRevLett.115.163601} time crystals.
 
The open Dicke model hosts a continuous phase transition, where the atoms reorganize from interacting individually with the cavity mode to forming a macroscopic dipole moment that collectively interacts with the cavity mode \cite{HEPP1973360}. It can be experimentally implemented using a Bose-Einstein condensate (BEC) strongly coupled to a longitudinal mode of an optical Fabry-P\'{e}rot type cavity, which is pumped by a retro-reflected laser beam with frequency $\omega_p$ and strength $\epsilon_p$, aligned perpendicularly to the cavity mode. In this configuration, the phase transition is characterized by the atoms spontaneously self-organizing into one of two possible density-wave configurations above a critical pump strength $\epsilon_c$. These density gratings enable efficient Bragg scattering of light from the transverse pump field into the cavity and vice versa \cite{2016HamburgSuperradiance, Baumann_2010, PhysRevLett.107.140402}. In a many-body picture, this transition can be understood as the result of tuning the strength of a light-induced infinite-range interaction (LIRI) between the atoms. 

There are two disparate regimes of LIRI in an atom-cavity system. In a broad-band cavity with $\kappa$ much larger than the recoil frequency $\omega_\mathrm{rec}$, the light sector dynamics is associated with a far more rapid time scale as compared to that of the matter sector such that the LIRI practically acts \textit{instantaneously} (I-LIRI). In a recoil-resolving narrow band cavity, where $\kappa$ is on the order of $\omega_\mathrm{rec}$, the time scales for the evolution in the light and matter sectors are comparable and hence the LIRI among the atoms is significantly \textit{retarded} (R-LIRI regime). The excitation modes exhibit distinct polaritonic character, i.e., a composition of comparable admixtures of photonic and matter components, and higher-order excitations are strongly suppressed due to the narrow cavity bandwidth \cite{Keszler_2014, 2016HamburgSuperradiance}.

In the vicinity of a phase boundary, many-body systems can give rise to intriguing signatures in their excitation spectra. In this work, we investigate the two expected polaritonic excitation modes in the R-LIRI regime close to the phase boundary of the Dicke-Hepp-Lieb phase transition of the open Dicke model. We experimentally demonstrate an intriguing domain, where the lower-energetic of these modes, instead of showing an ordinary roton-type mode softening \cite{RotonExcSpectrum}, is found to approach zero energy at a pump strength $\epsilon_z$ well below the critical pump strength $\epsilon_c$ at the Dicke-Hepp-Lieb transition boundary. The excitation energy remains zero across a finite pump strength interval $[\epsilon_z,\epsilon_c]$, such that a peculiar situation arises, where an excitation is possible at zero energy cost, but nevertheless no instability of the system is created.

Indications of this phenomenon have been previously found in calculations for a broad-band cavity I-LIRI scenario \cite{Oez:2012} with the result of a tiny zero-energy pump strength interval, more than six orders of magnitude smaller than the critical pump strength ($1-\frac{\epsilon_z}{\epsilon_c} \sim 10^{-6}$) and as such practically not observable. In fact, experimentally, in Ref.\cite{RotonExcSpectrum}, a single excitation mode with an ordinary roton-like mode softening with $\epsilon_z \sim \epsilon_c$ was found with no discernable zero-energy domain. In a complementary theory work \cite{Nag:2011} for a narrow-band R-LIRI cavity system, a similar zero-energy excitation mode was found with zero energy attained across a much larger pump strength interval only one order of magnitude smaller than the critical pump strength. In a more recent review article \cite{Mivehvar_2021}, the authors explicitly speculated that the domain of zero-energy excitations might be in reach in a R-LIRI cavity system. The lack of an experimental confirmation of such excitations, more than one decade after their mention by theory, reflects the extensive technical complexity to realize the R-LIRI scenario necessary for a sizable experimentally observable effect. See the supplemental material \cite{suppmat} for a discussion of the size of the zero-energy interval $[\epsilon_z,\epsilon_c]$ for I-LIRI and R-LIRI regimes. 

In our experiment, a BEC of $N_\mathrm{a}\approx40\times10^3$ \textsuperscript{87}Rb atoms is strongly coupled to a single mode of an optical cavity with a finesse of $4.2\times 10^5$. The atoms are primarily addressed with a retro-reflected laser beam with frequency $\textit{$\omega$}_\mathrm{p}$, transversely oriented with respect to the cavity axis, referred to in the following as the transverse pump. This setup is depicted in Fig.~\ref{fig:1}(a). The pump wavelength $\textit{$\lambda$}_\mathrm{p}=803.26$ \si{\nano m} is far-red detuned with respect to the relevant atomic transitions, namely the $D_\mathrm{1}$ and $D_\mathrm{2}$ lines at $794.97$ and $780.24$ \si{\nano m}, respectively, thus we operate in the dispersive regime. At this chosen pump wavelength, the transverse pump produces an attractive standing-wave potential at the position of the atoms. In this experiment, we utilise a cavity that operates in the recoil-resolved regime, where the field decay rate of $\textit{$\kappa$}=2\textit{$\pi$}\times 3.60$ \si{\kilo Hz} is comparable to the recoil frequency $\textit{$\omega$}_\mathrm{rec}=2\textit{$\pi$} \times 3.55$ \si{\kilo Hz} \cite{Keszler_2014}. The cavity resonance is shifted as a result of the refractive index of the BEC in the cavity by an amount $\textit{$\delta$}_- = N_\mathrm{a} U_\mathrm{0}/2$, where $U_\mathrm{0} = - 2\textit{$\pi$}\times 0.34$ \si{Hz} is the dispersive shift per atom. We therefore define an effective detuning with respect to the cavity resonance of $\textit{$\delta$}_\mathrm{eff}=\textit{$\delta$}_\mathrm{c}-\textit{$\delta$}_-$, where $\textit{$\delta$}_\mathrm{c}=\textit{$\omega$}_\mathrm{p}-\textit{$\omega$}_\mathrm{c}$ is the detuning of the transverse pump frequency $\textit{$\omega$}_\mathrm{p}$ with respect to the empty cavity resonance frequency $\textit{$\omega$}_\mathrm{c}$. In this experimental platform, the two-level systems at the basis of the open Dicke model are atomic momentum modes, with the zero momentum mode $\ket{\mathrm{p_y},\mathrm{p_z}}\equiv\ket{0,0}$, referred to as the BEC mode, forming the ground state, and a superposition $\ket{\pm1,\pm1} \equiv \frac{1}{2}\sum_{\nu,\mu \in \{-1,+1\}} \ket{\nu \hbar k,\mu \hbar k}$ of $\ket{\pm \hbar k,\pm \hbar k}$ momentum modes forming the excited state. Here, the same wavenumber $k=2\textit{$\pi$}/\textit{$\lambda$}_\mathrm{p}$ is assumed for both the transverse pump and the longitudinal probe waves, which is justified due to their comparatively small detunings on the scale of a few \si{\kilo Hz}. When the transverse pump is red-detuned with respect to the cavity resonance ($\textit{$\delta$}_\mathrm{eff}<0$) and its pump strength $\textit{$\epsilon$}_\mathrm{p}$ exceeds a critical value $\textit{$\epsilon$}_\mathrm{c}$, the atomic ensemble transists from the normal phase (NP), a BEC trapped in the standing wave potential of the pump, into one of two density lattice configurations available in the superradiant (SR) phase \cite{PhysRevLett.107.140402, Dissipativetimecrystal}, resulting in the build-up of a finite intracavity light field. By scanning the \{$\textit{$\epsilon$}_\mathrm{p}$, $\textit{$\delta$}_\mathrm{eff}$\}-plane and measuring the number $N_\mathrm{p}$ of photons leaking out of the cavity with a heterodyne detector, we obtain the phase diagram shown in Fig.~\ref{fig:1}(c). It should be noted that the critical pump strength $\textit{$\epsilon$}_\mathrm{c}$ varies with $\textit{$\delta$}_\mathrm{eff}$.

\begin{figure}[ht]
\centering
\includegraphics[width=\linewidth]{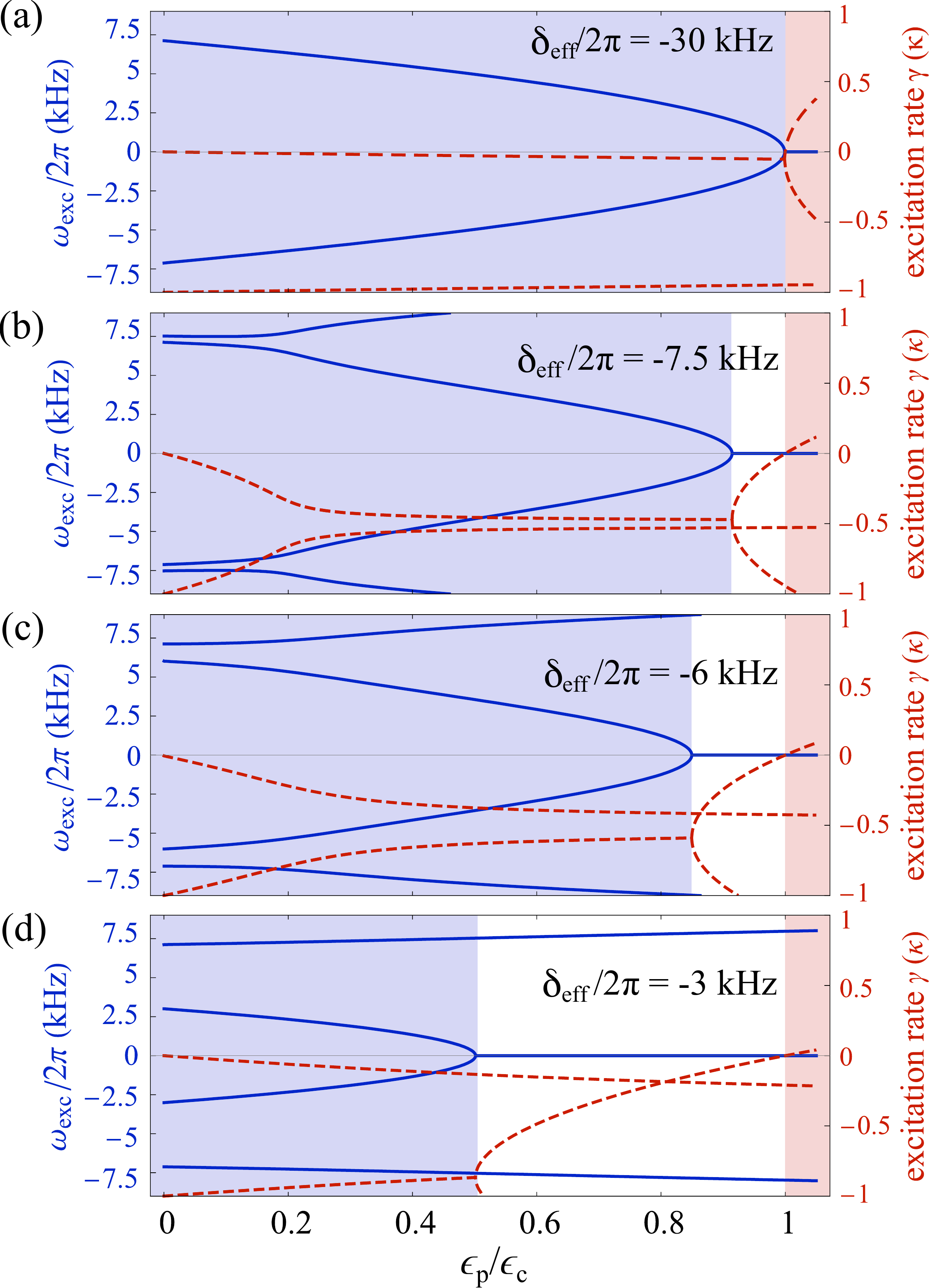}
\caption{Calculated excitation frequencies (blue, solid lines) and excitation rates (red, dashed lines) of the lower and upper polariton modes for different effective detunings $\textit{$\delta$}_\mathrm{eff}$ in the NP ($\epsilon_{\textrm{p}}<\epsilon_{\textrm{c}}$). In the red area, where $\epsilon_{\textrm{p}}>\epsilon_{\textrm{c}}$, a correct linearization analysis should employ a steady state within the SR phase. (a): The lower polariton mode at zero pump strength shows an excitation frequency of $\pm \omega_{\textrm{rec}} / \pi = \pm 7.1\,$kHz, which identifies its atomic character for $\epsilon_{\textrm{p}}=0$. As $\epsilon_{\textrm{p}}$ increases, its frequency falls to zero very close to the critical pump strength $\epsilon_{\textrm{c}}$. The upper polariton mode resides at excitation frequencies around $\pm \delta_{\textrm{eff}} / 2\pi = \pm 30\,$kHz outside the plot range for all values of $\epsilon_{\textrm{p}}$, and hence identifies its photonic character. The excitation rates are negative nearly all the way to $\epsilon_{\textrm{p}} = \epsilon_{\textrm{c}}$.(b): As in (a), at $\epsilon_{\textrm{p}}=0$ the upper mode and the lower mode show purely photonic and atomic character, respectively. The lower polariton mode starts with an excitation frequency $\pm \omega_{\textrm{rec}} / \pi$ at $\epsilon_{\textrm{p}}=0$ as in (a) but clearly falls to zero before the critical pump strength is reached. (c) and (d): On a large part of the pump strength axis (white area), zero excitation frequency is found for the lower polariton, while all excitation rates are negative below the critical pump strength.}
\label{fig:2}
\end{figure}

To theoretically analyse the system as it approaches the phase transition, we study the excitation energy of the polaritonic modes by performing a stability analysis. In this model, we only take into account the occupation of the two lowest momentum modes, $\ket{0,0}$ and $\ket{\pm1,\pm1}$, which we refer to as the two-mode stability analysis \cite{doi:10.1073/pnas.1417132112}. We perform this analysis for a fixed $\textit{$\delta$}_\mathrm{eff}$ to determine the excitation energy $E_\mathrm{exc}=\hbar\,\textit{$\omega$}_\mathrm{exc}$ and excitation rate $\gamma$ of the polariton modes, resulting from the real and imaginary parts of the eigenvalues of the stability matrix, respectively. Further details on the stability analysis and the stability matrix itself are provided in the supplemental material \cite{suppmat}. For zero transverse pump strength $\epsilon_\mathrm{p}=0$, the atomic and photonic modes are decoupled, the energy required to excite the atoms from the $\ket{0,0}$ into the $\ket{\pm1,\pm1}$ momentum modes is $2\hbar\,\textit{$\omega$}_\mathrm{rec}$, and the energy required to excite the cavity mode is $|\hbar\textit{$\delta$}_\mathrm{eff}|$, see Fig.~ \ref{fig:1}(b). As the transverse pump strength increases, the off-diagonal terms in the stability matrix become nonzero, and the stability matrix obtains two new eigenstates, composed of mixed atom-light excitations, referred to as the upper and lower polariton modes. The excitation energy of the lower polariton mode reaching zero is a necessary but not sufficient requirement for spontaneously transitioning into the SR phase. The second necessary criterion is a positive excitation rate, as this determines whether small deviations from the equilibrium are self-reinforcing. In the case of a positive excitation rate, the system is pulled into the SR phase. For a negative excitation rate, a transition to the SR phase is damped, causing the system to remain in the NP.

The results of the two-mode stability analysis are shown in Figs.~\ref{fig:2}(a-d). In the far detuned  case $|\textit{$\delta$}_\mathrm{eff}| \gg \omega_\mathrm{rec}$, approximately realized in Fig.~\ref{fig:2}(a), the lower polariton mode contains a very small admixture of the photonic component, and its excitation energy undergoes a roton-like mode softening \cite{RotonExcSpectrum}, 
\begin{equation}
\textit{$\omega$}_\mathrm{exc}=2\,\textit{$\omega$}_\mathrm{rec}\,\sqrt{1-\frac{\textit{$\epsilon$}_\mathrm{p}}{\textit{$\epsilon$}_\mathrm{c}}},
\label{eq:sqrt}
\end{equation}
with the excitation energy reaching zero at a pump strength nearly coinciding with the critical pump strength $\textit{$\epsilon$}_\mathrm{c}$. As seen in Fig.~\ref{fig:2}(a) (red graph), the associated excitation rate $\textit{$\gamma$}$ is initially drawn to negative values before splitting into two branches very near to the critical pump strength, with the upper branch becoming positive exactly at the critical pump strength $\textit{$\epsilon$}_\mathrm{c}$. This behavior resembles aforementioned previous observations in the I-LIRI regime \cite{RotonExcSpectrum, Landig_2015}, because $\kappa \gg \omega_\mathrm{rec}$ and $|\delta_\mathrm{eff}| \gtrsim \kappa$ such that the condition of far-detuning $|\delta_\mathrm{eff}| \gg \omega_\mathrm{rec}$ is generally fulfilled.

In the near-resonant regime $|\delta_\mathrm{eff}| \lesssim 2\, \omega_\mathrm{rec}$, Fig.~\ref{fig:2}(b-d), the admixture of the photonic component to the polariton mode becomes significantly larger as the characteristic photon energy $\hbar |\delta_\mathrm{eff}|$ approaches the relevant atomic energy $2\,\hbar \omega_\mathrm{rec}$. While decay of the atomic momentum modes is negligible on experimental timescales, the photons decay from the cavity at a rate of $1/(2\textit{$\pi$}\textit{$\kappa$})$. Analytical calculations using the Dicke model for the resonant case $|\delta_\mathrm{eff}| = 2\,\omega_\mathrm{rec}$ shown in the supplemental material \cite{suppmat} suggest an interpretation that the dissipation effectively shifts the excitation rate curve by $-\kappa/2$, which means that the critical pump strength is pushed to higher values than in the case for $\kappa=0$. This is consistent with previous results for the polariton frequencies \cite{Dimer2007, Jara2024}. This effect becomes more pronounced as $|\delta_\mathrm{eff}|$ decreases and leads to a regime in which the lower polaritonic excitation frequency $\omega_\mathrm{exc}$ reaches zero at a value $\epsilon_\mathrm{z}$ significantly below $\epsilon_\mathrm{c}$. However, due to the dissipation-induced overdamping, $\gamma < 0$, the system remains in the homogeneous NP despite the vanishing energy cost for the excitation of the lower polariton mode. We refer to this mode as a \textit{polaritonic zero-energy excitation mode}.

\begin{figure}[h]
\centering
\includegraphics[width=\linewidth]{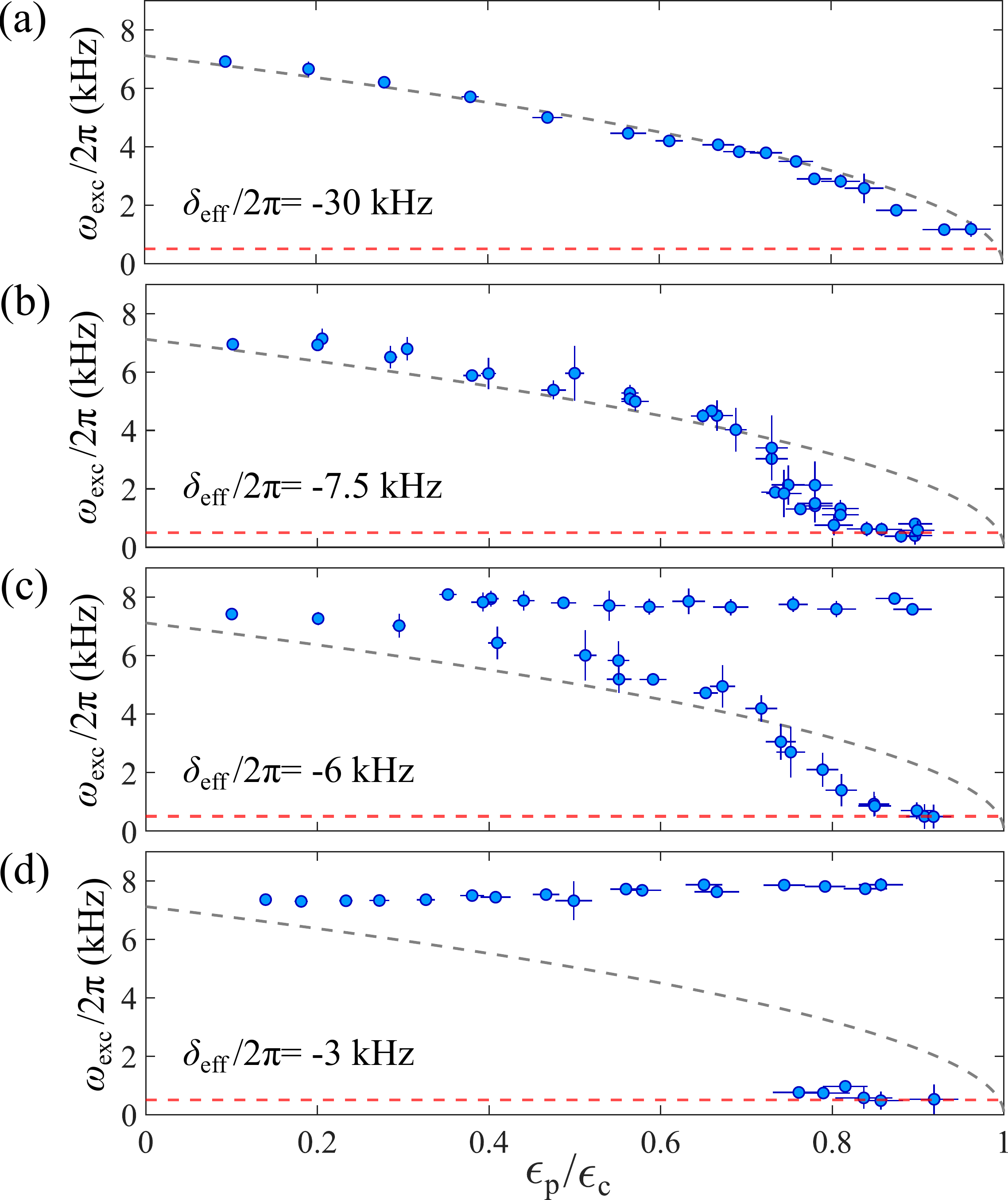}
\caption{Excitation spectra measured for different effective detunings $\textit{$\delta$}_\mathrm{eff}$ according to Fig.~\ref{fig:2}. The lowest frequency that can be accurately resolved with our experimental protocol, $500$ \si{Hz}, is marked by a red dashed line, and the square-root behavior predicted for the far-detuned regime as per equation Eq.~(\ref{eq:sqrt}) is shown in grey. (a) shows the mode softening of the lower polariton mode as the transverse pump strength $\textit{$\epsilon$}_\mathrm{p}$ approaches $\textit{$\epsilon$}_\mathrm{c}$. (b) shows the formation of an overdamped excitation mode, as the excitation energy of the lower polariton mode falls off to zero and remains at zero before $\textit{$\epsilon$}_\mathrm{p}$ reaches the critical pump strength $\textit{$\epsilon$}_\mathrm{c}$. In (c) and (d), for $\epsilon_{\textrm{p}}=0$, the upper (lower) polariton mode acquires pure atomic (photonic) character. Away from $\epsilon_{\textrm{p}}=0$, both modes possess sufficient atomic admixture to be observable with our detection protocol. Horizontal error bars represent the uncertainty of the critical pump strength. The vertical error bars represent the standard deviation of the excitation energies around the mean as well as the respective uncertainties from the data evaluation method.}
\label{fig:3}
\end{figure}

We record the excitation energy $\hbar \omega_\mathrm{exc}$ of the polariton modes using a form of Bragg spectroscopy \cite{Stenger_1999, RotonExcSpectrum, Ernst_2009}. We therefore utilize an additional beam superimposed on the cavity axis (the $z$-axis), in addition to the previously-introduced transverse pump along the $y$-axis. This additional beam is referred to in the following as the longitudinal probe with frequency $\omega_\mathrm{l}$, see Fig.~\ref{fig:1}(a). When the frequency detuning $\textit{$\delta$}_\mathrm{lt}$ between the transverse pump and the longitudinal probe is equal to the energy required to excite one of the polariton modes, the atoms can scatter light between pump and probe, depending on the relative detuning, and create excitations of the polariton modes, as illustrated in Figs.~\ref{fig:1}(a) and \ref{fig:1}(b). As the atomic component of the polariton modes is a superposition of the $\ket{\pm1,\pm1}$ momentum modes, this results in the build-up of a weak density grating, which fulfills the Bragg condition for constructive interference of scattering from the transverse pump into the cavity and vice versa. In this case, this scattering process is temporary, as the modes are subject to damping, and the process is not self-reinforcing below $\textit{$\epsilon$}_\mathrm{c}$. Thus, Bragg spectroscopy can be used to determine the excitation energy of the momentum modes by determining the detuning $\textit{$\delta$}_\mathrm{lt}$ at which most of the atoms are excited into the polariton mode, and correspondingly, most transverse pump light is scattered into the cavity. The highest intensity of light scattered into the cavity occurs when the detuning matches the corresponding excitation energy, $\delta_\mathrm{lt} = \hbar \omega_\mathrm{exc}$. The results are presented in Fig.~\ref{fig:3}. The detailed protocol for how we extract $\textit{$\omega$}_\mathrm{exc}$ from the measured intracavity photon number $N_\mathrm{p}$ can be found in the supplementary material \cite{suppmat}.

In the far-detuned regime $|\textit{$\delta$}_\mathrm{eff}| \gg \omega_\mathrm{rec}$, at an effective detuning of $\textit{$\delta$}_\mathrm{eff}/2\textit{$\pi$}=-30$ \si{\kilo Hz}, we obtain the excitation spectrum shown in Fig.~\ref{fig:3}(a). The excitation energy of the lower polariton mode falls off in good agreement with Eq.~\ref{eq:sqrt}, which is also captured by our two-mode stability analysis shown in Fig.~\ref{fig:2}(a). This shows the expected behavior for a small admixture of the photonic component to the lower polariton mode, with the excitation rate falling off to zero close to $\textit{$\epsilon$}_\mathrm{c}$. The small deviation close to $\textit{$\epsilon$}_\mathrm{c}$ as compared to the results for a significantly larger $\textit{$\kappa$}$ \cite{RotonExcSpectrum} is attributed to a larger admixture of the photonic component to the lowest polariton mode in our system, which could contribute additional damping. This is further supported by our excitation spectrum measurements for an effective detuning of $\textit{$\delta$}_\mathrm{eff}/2\textit{$\pi$}=-15$ \si{\kilo Hz} (see supplementary material \cite{suppmat}), which are closer to the cavity resonance and differ more from the behavior predicted by Eq.~\ref{eq:sqrt} for the $|\textit{$\delta$}_\mathrm{eff}| \gg \omega_\mathrm{rec}$ regime.

When the transverse pump is near-resonant, $|\delta_\mathrm{eff}| \lesssim 2\, \omega_\mathrm{rec}$, we see significant deviations from the prediction of Eq.~\ref{eq:sqrt}. The energetically lowest polariton mode displays mode softening, however, such that the excitation energy reaches zero at a value $\textit{$\epsilon$}_\mathrm{z}$ notably below $\textit{$\epsilon$}_\mathrm{c}$. This can be clearly seen in Fig.~\ref{fig:3}(b), where the excitation frequency falls close to zero at $\textit{$\epsilon$}_\mathrm{z}/\textit{$\epsilon$}_\mathrm{c}\approx0.8$. Due to the digital bandpass filter we applied to the measured photon number, we cannot experimentally resolve frequencies below $500$ \si{Hz}, indicated by red dashed lines in Figs.~\ref{fig:3}(a-d). The results of the excitation frequency tending towards zero are also confirmed theoretically by the two-mode stability analysis and the dynamical multi-mode simulations (see supplementary material \cite{suppmat}). This observation verifies the presence of a polaritonic zero-energy excitation mode, which is accessible due to the recoil resolving ultralow bandwidth of our optical cavity.

Since the fundamental cavity mode excitation in the absence of a pump is $\hbar \delta_\mathrm{eff}$, it is interesting to explore the case when the cavity photon excitation energy is slightly below the atomic excitation energy, which is the case for $\delta_\mathrm{eff}/2\pi =-6$ \si{\kilo Hz}. This results in a strong mixture of atomic and photonic components for both upper and lower polaritonic modes as the transverse pump strength increases. In Fig.~\ref{fig:3}(c), we observe two polaritonic modes with different excitation energies that deviate significantly from each other beyond $\textit{$\epsilon$}_\mathrm{p}/\textit{$\epsilon$}_\mathrm{c}=0.3$ with one slightly increasing and the other tending towards zero at $\textit{$\epsilon$}_\mathrm{z}/\textit{$\epsilon$}_\mathrm{c}\approx0.80$. We interpret this result as the observation of both upper and lower polaritonic modes. While the upper polaritonic mode was also predicted for $\textit{$\delta$}_\mathrm{eff}/2\textit{$\pi$}=-7.5$ \si{\kilo Hz} in Fig.~\ref{fig:2}(b), we were unable to verify it experimentally because our measurement protocol is primarily sensitive to polaritonic modes with a significant atomic character. Therefore, the detection of both the upper and lower polariton modes throughout the excitation spectrum confirms their strong mixtures of atomic and photonic sectors. This is further corroborated by the absence of a multi-peak structure in the excitation curves from the dynamical multi-mode simulations, which are presented in the supplementary material \cite{suppmat}.

For smaller detunings, $|\textit{$\delta$}_\mathrm{eff}| < \textit{$2\,\omega$}_\mathrm{rec}$, for example, at $\textit{$\delta$}_\mathrm{eff}/2\textit{$\pi$}=-3$ \si{\kilo Hz}, the deviations from the prediction of Eq.~\ref{eq:sqrt} become even more pronounced, as is shown in Fig.~\ref{fig:3}(d). Here, we observe an upper polariton mode with an excitation frequency that begins at $\textit{$\omega$}_\mathrm{exc}\approx\textit{$2\,\omega$}_\mathrm{rec}$ at $\textit{$\epsilon$}_\mathrm{p}=0$, and increases from there to around $8$ \si{\kilo Hz}. At a transverse pump strength of $\textit{$\epsilon$}_\mathrm{p}/\textit{$\epsilon$}_\mathrm{c}\geq0.75$, a second mode appears with an excitation frequency that remains around zero and elicits a significantly stronger response than the upper polariton mode, which significantly weakens past this point for the entire range of parameters, where the lower polariton mode can be detected. Below this pump strength, this mode, which is predicted in the two-mode stability analysis in Fig.~\ref{fig:2}(d), is not observable with the applied experimental protocol. While this may provide some indication as to the evolution of the atomic and photonic components of the polaritonic modes with $\textit{$\epsilon$}_\mathrm{p}/\textit{$\epsilon$}_\mathrm{c}$ in the near-resonant regime, more extensive simulations, including a full eigenfunction analysis, would be required to confirm this, which goes beyond the scope of this work.

In conclusion, in the normal phase of the open Dicke model, implemented in an atom cavity platform, we experimentally observed a polaritonic zero-energy excitation mode, for which, despite vanishing excitation energy, no instability of the normal phase arises. Theoretical analysis confirms that the weak dissipation in the photonic sector, with the intracavity field decay rate being smaller than the photon recoil frequency, is responsible for the observations. We speculate that our observations may prove advantageous for critical quantum metrology applications \cite{Gie:2022, Hot:2024}, since an infinitely steep excitation energy change becomes accessible at $\epsilon_{\textrm{z}}$, well separated from the critical point $\epsilon_{\textrm{c}}$ where a phase transition occurs.

We thank J. Klinner, M. Wolke, J. Skulte, and L. Mathey for their support during the early stages of the project. We thank T. Donner, J. Marino, S. B. J{\" a}ger, and C. Zimmermann for fruitful discussions. This work was funded by the QuantERA II Program that has received funding from the European Union's Horizon 2020 research and innovation program under Grant Agreement No. 101017733, the Deutsche Forschungsgemeinschaft (DFG, German Research Foundation) ``SFB-925" project 170620586, and the Cluster of Excellence ``Advanced Imaging of Matter" (EXC 2056), Project No. 390715994. J.G.C. acknowledges financial support from the National Academy of Science and Technology Philippines (NAST PHL).

\bibliography{references.bib}

\end{document}



\title{Supplemental Materials for \\  Observation of a zero-energy excitation mode in the open Dicke model}

\author{Anton B\"{o}lian}
\thanks{These authors have contributed equally to this work.}
\affiliation{Institute for Quantum Physics, Universit\"at Hamburg, 22761 Hamburg, Germany}

\author{Phatthamon Kongkhambut}
\thanks{These authors have contributed equally to this work.}
\affiliation{Institute for Quantum Physics, Universit\"at Hamburg, 22761 Hamburg, Germany}
\affiliation{Quantum Simulation Research Laboratory, Department of Physics and Materials Science, Faculty of Science, Chiang Mai University, Chiang Mai, 50200, Thailand}
\affiliation{Thailand Center of Excellence in Physics, Office of the Permanent Secretary, Ministry of Higher Education, Science, Research and Innovation, Thailand}

\author{Christoph Georges}
\affiliation{Institute for Quantum Physics, Universit\"at Hamburg, 22761 Hamburg, Germany}

\author{Roy D. Jara Jr.}
\affiliation{National Institute of Physics, University of the Philippines, Diliman, Quezon City 1101, Philippines}

\author{Jos\'{e} Vargas}
\affiliation{Institute for Quantum Physics, Universit\"at Hamburg, 22761 Hamburg, Germany}

\author{Jens Klinder}
\affiliation{Institute for Quantum Physics, Universit\"at Hamburg, 22761 Hamburg, Germany}

\author{Jayson G. Cosme}
\affiliation{National Institute of Physics, University of the Philippines, Diliman, Quezon City 1101, Philippines}

\author{Hans Ke{\ss}ler}
\email[]{hkessler@physnet.uni-hamburg.de}
\affiliation{Institute for Quantum Physics, Universit\"at Hamburg, 22761 Hamburg, Germany}

\author{Andreas Hemmerich}
\email[]{andreas.hemmerich@uni-hamburg.de}
\affiliation{Institute for Quantum Physics, Universit\"at Hamburg, 22761 Hamburg, Germany}

\date{\today}

\maketitle

\onecolumngrid
\newpage

\section{Atom-cavity model}
In the following, we will analyze the atom-cavity Hamiltonian, going through it term-by-term and discussing these individually. We restrict our analysis to two dimensions, as the transverse pump and intracavity light field run along these directions, and almost all the dynamics in the system take place within this plane. In this approximation, the full Hamiltonian is as follows:
\begin{equation}
    \hat{H}_\mathrm{full}=\hat{H}_\mathrm{c}+\hat{H}_{a}+\hat{H}_\mathrm{aa}+\hat{H}_\mathrm{int}
\end{equation}
The cavity term is described by the following equation,
\begin{equation}
    \hat{H}_\mathrm{c}=-\hbar\textit{$\delta$}_\mathrm{c} \hat{a}^\dagger \hat{a},
\end{equation}
wherein $\textit{$\delta$}_\mathrm{c}$ is the bare detuning between the transverse pump and the cavity resonance, and $\hat{a}$ is the annihilation operator that annihilates a single photon in the cavity mode. The purely atomic term is given by
\begin{equation}
    \hat{H}_\mathrm{a}=\int\hat{\textit{$\Psi$}}^\dagger(y,z)\left(-\frac{\hbar^2\nabla^2}{2m} +\textit{$\epsilon$}_\mathrm{p}\cos^2(ky)\right)\hat{\textit{$\Psi$}}(y,z)\ dydz,
\end{equation}
where $m$ is the mass of a rubidium $87$ atom, $\textit{$\epsilon$}_\mathrm{p}$ is the transverse pump strength, and $k$ is the wave vector associated with the transverse pump wavelength $\textit{$\lambda$}_\mathrm{p}$. Here, on the right hand side of the equation, we neglect the harmonic magnetic trap potential, present in the experiment. This is justified since the magnetic forces, which act to hold the atom sample in place, are much smaller than any other forces in the atom-cavity dynamics. Their neglect amount to realizing the thermodynamic limit. The on-site interaction between the atoms is captured by
\begin{equation}
    \hat{H}_\mathrm{aa}=U_\mathrm{a}\int\hat{\textit{$\Psi$}}^\dagger(y,z)\hat{\textit{$\Psi$}}^\dagger(y,z)\hat{\textit{$\Psi$}}(y,z)\hat{\textit{$\Psi$}}(y,z)\ dydz.
\end{equation}
In this equation, $U_\mathrm{a}=\sqrt{2\textit{$\pi$}}a_\mathrm{s}\hbar/ml_\mathrm{x}$ is the effective two-dimensional scattering length along the $\mathrm{y-z}$ directions and $l_\mathrm{x}$ is the harmonic oscillator length of the magnetic trap along the $x$ direction. For the sake of simplicity and because of the expectation that contact interaction will not critically alter the polaritonic excitation spectra at low atomic densities, we henceforth neglect $U_\mathrm{a}$. The interaction between the atoms and the cavity is described by the following term:
\begin{equation}
    \hat{H}_\mathrm{int}=\int\hat{\textit{$\Psi$}}^\dagger(y,z)\left(\hbar U_\mathrm{0}\cos^2(kz)\hat{a}^\dagger\hat{a}+\sqrt{\hbar\textit{$\epsilon$}_\mathrm{p} U_\mathrm{0}}\cos(ky)\cos(kz)\left(\hat{a}^\dagger+\hat{a}\right) \right)\hat{\textit{$\Psi$}}(y,z)\ dydz,
\end{equation}
where $U_\mathrm{0}$ is the dispersive shift per intracavity photon. To simulate our system below the critical threshold, we perform a mean-field approximation, replacing the operators $\hat{a}$ with their expectation values $\alpha$, and obtaining new equations of motion, which are as follows, starting with the matter component:
\begin{equation}
\begin{aligned}
    i\hbar\frac{\partial}{\partial t}\textit{$\Psi$}(y,z,t)&=\biggl(-\frac{\hbar^2\nabla^2}{2m}+\hbar U_\mathrm{0}|\textit{$\alpha$}(t)|^2\cos^2(kz)+\hbar U_\mathrm{0}|\textit{$\alpha$}_\mathrm{p}(t)|^2\cos^2(ky)\\
    &+2\hbar U_\mathrm{0} \mathrm{Re}(\textit{$\alpha$}(t))\,\textit{$\alpha$}_\mathrm{p}(t)\cos(kz)\cos(ky)\biggr)\textit{$\Psi$}(y,z,t),
\end{aligned}
\end{equation}
where $\textit{$\alpha$}_\mathrm{p}(t) \equiv \sqrt{\epsilon_\mathrm{p}(t)/\hbar U_0}$ denotes the nondimensionalized amplitude of the pump field. Meanwhile, the intracavity field is described by the following terms,
\begin{equation}
\begin{aligned}
    i\hbar\frac{\partial}{\partial t}\textit{$\alpha$}(t) =\hbar\left(-\textit{$\delta$}_\mathrm{c}+N_\mathrm{a}U_\mathrm{0}\langle\cos^2(kz)\rangle-i\textit{$\kappa$} \right)\textit{$\alpha$}(t)+\hbar N_\mathrm{a}U_\mathrm{0}\,\textit{$\alpha$}_\mathrm{p}(t)\langle\cos(ky)\cos(kz) \rangle,
\end{aligned}
\end{equation}
wherein the expectation values $\langle...\rangle$ are integrated over the volume of the BEC, weighted by its density. Following this approximation, we perform a plane-wave expansion of the atomic wave function in the relevant $y-z$ plane:
\begin{equation}
    \textit{$\Psi$}(y,z,t)=\sum_\mathrm{n,m}\textit{$\phi$}_\mathrm{n,m}e^{inky}e^{imkz}.
\end{equation}
In this equation, $\textit{$\phi$}_\mathrm{n,m}$ are the normalised $(\sum_\mathrm{n,m}|\textit{$\phi$}_\mathrm{n,m}|^2=1)$ single-particle amplitudes of the momentum state $(p_y,p_z)=(m,n)\hbar k$. This leads to the following equation of motion for the atomic density:
\begin{equation}
\begin{aligned}
    i\frac{\partial}{\partial t}\textit{$\phi$}_\mathrm{n,m}&=\left(\textit{$\omega$}_\mathrm{rec}(n^2+m^2)+\frac{U_\mathrm{0}|\textit{$\alpha$}|^2}{2}+\frac{U_\mathrm{0}|\textit{$\alpha$}_\mathrm{p}|^2}{2}\right)\textit{$\phi$}_\mathrm{n,m}+\frac{U_\mathrm{0}}{4}|\textit{$\alpha$}(t)|^2(\textit{$\phi$}_\mathrm{n,m+2}+\textit{$\phi$}_\mathrm{n,m-2})+\frac{U_\mathrm{0}}{4}|\textit{$\alpha$}_\mathrm{p}(t)|^2(\textit{$\phi$}_\mathrm{n+2,m}+\textit{$\phi$}_\mathrm{n-2,m})\\
    &+\frac{U_\mathrm{0}}{4}\textit{$\alpha$}_\mathrm{p}(t)Re(\textit{$\alpha$}(t))\left(\textit{$\phi$}_\mathrm{n+1,m+1}+\textit{$\phi$}_\mathrm{n+1,m-1}+\textit{$\phi$}_\mathrm{n-1,m+1}+\textit{$\phi$}_\mathrm{n-1,m-1} \right),
\end{aligned}
\label{EOMmatterfull}
\end{equation}
and the following equation of motion for the cavity mode:
\begin{equation}
\begin{aligned}
    i \frac{\partial}{\partial t}\textit{$\alpha$}&=\left(-\textit{$\delta$}_\mathrm{c}+\frac{N_\mathrm{a}U_\mathrm{0}}{2}\left(\sum_\mathrm{n,m}Re\left(\textit{$\phi$}_\mathrm{n,m}\textit{$\phi$}^*_\mathrm{n,m+2} \right)-1\right)-i\textit{$\kappa$}\right)\textit{$\alpha$}\\ 
    &-i\frac{U_\mathrm{0}N_\mathrm{a}}{4}\textit{$\alpha$}_\mathrm{p}(t)\left(\sum_\mathrm{n,m}\textit{$\phi$}_\mathrm{n,m}\left(\textit{$\phi$}_\mathrm{n+1,m+1}^*+\textit{$\phi$}_\mathrm{n-1,m+1}^*\right)+\textit{$\phi$}_\mathrm{n,m}^*\left(\textit{$\phi$}_\mathrm{n+1,m+1}+\textit{$\phi$}_\mathrm{n-1,m+1}\right)\right).
\end{aligned}
\label{EOMlightfull}
\end{equation}
When the system is in the normal phase, below the self-organisation threshold, we can restrict our analysis to the $\ket{0,0}$ and a superposition of the $\ket{\pm1,\pm1}$ momentum modes, with $\textit{$\phi$}_{1}=\frac{1}{2}\left(\textit{$\phi$}_{1,1}+\textit{$\phi$}_{1,-1}+\textit{$\phi$}_{-1,1}+\textit{$\phi$}_{-1,-1}\right)$, which is justified below the critical threshold, as higher momentum modes do not play a significant role below $\textit{$\epsilon$}_\mathrm{c}$. In the following, we rescale the intracavity amplitude to $|\textit{$\beta$}|^2=|\textit{$\alpha$}|^2U_\mathrm{0}/4\textit{$\omega$}_\mathrm{rec}$ and the pump strength is rescaled to $\textit{$\epsilon$}_\mathrm{p}=|\textit{$\alpha$}_\mathrm{p}|^2U_\mathrm{0}/\textit{$\omega$}_\mathrm{rec}$. Furthermore, we can reasonably approximate our state below the critical threshold as $\textit{$\phi$}_\mathrm{n,m}=\textit{$\delta$}_\mathrm{n,0}\textit{$\delta$}_{m,0}$ and $\textit{$\beta$}=0$, i.e., $100\%$ condensate occupation and no photons in the cavity mode. We then linearise the equations of motion \ref{EOMmatterfull} and \ref{EOMlightfull} around this equilibrium state, and set the energy of $\ket{0,0}$ to the zero point energy, allowing us to write these four linear equations as follows \cite{CGeorgesPhDThesis}:
\begin{equation}\label{eq:stab_mat}
i\frac{\partial}{\partial t}
\begin{pmatrix}
    \textit{$\beta$}\\
    \textit{$\beta$}^*\\
    \textit{$\phi$}_\mathrm{1}\\
    \textit{$\phi$}_\mathrm{1}^*\\
\end{pmatrix}
=
\begin{pmatrix}
    \textit{$\delta$}_\mathrm{eff}-i\textit{$\kappa$} & 0 & -i\frac{N_\mathrm{a}U_\mathrm{0}}{4}\sqrt{\textit{$\epsilon$}_\mathrm{p}} & -i\frac{N_\mathrm{a}U_\mathrm{0}}{4}\sqrt{\textit{$\epsilon$}_\mathrm{p}}\\
    0 & -\textit{$\delta$}_\mathrm{eff}-i\textit{$\kappa$} &  -i\frac{N_\mathrm{a}U_\mathrm{0}}{4}\sqrt{\textit{$\epsilon$}_\mathrm{p}} & -i\frac{N_\mathrm{a}U_\mathrm{0}}{4}\sqrt{\textit{$\epsilon$}_\mathrm{p}}\\
    i\textit{$\omega$}_\mathrm{rec}\sqrt{\textit{$\epsilon$}_\mathrm{p}}&-i\textit{$\omega$}_\mathrm{rec}\sqrt{\textit{$\epsilon$}_\mathrm{p}}&2\textit{$\omega$}_\mathrm{rec}&0 \\
    -i\textit{$\omega$}_\mathrm{rec}\sqrt{\textit{$\epsilon$}_\mathrm{p}}&i\textit{$\omega$}_\mathrm{rec}\sqrt{\textit{$\epsilon$}_\mathrm{p}}&0&-2\textit{$\omega$}_\mathrm{rec}\\
\end{pmatrix}
\begin{pmatrix}
    \textit{$\beta$}\\
    \textit{$\beta$}^*\\
    \textit{$\phi$}_\mathrm{1}\\
    \textit{$\phi$}_\mathrm{1}^*\\
\end{pmatrix}
\end{equation}
By diagonalising this stability matrix and calculating its eigenvalues, one can separate out the real and imaginary parts of the eigenvalues, which correspond to the excitation energy $E_\mathrm{exc}=\hbar\textit{$\omega$}_\mathrm{exc}$ and excitation rate $\textit{$\gamma$}$ of the mode. For $\textit{$\epsilon$}_\mathrm{p}=0$, the modes are fully decoupled, and the eigenstates of the matrix are excitations of $\textit{$\phi$}_\mathrm{1}$ and $\textit{$\beta$}$, i.e., the atomic momentum state and the cavity mode. For nonzero $\textit{$\epsilon$}_\mathrm{p}$, the eigenstates of this matrix are mixed atom-light excitations, or polariton modes. The degree that these modes mix depends on the detuning between the transverse pump and the cavity and on its relation to the cavity decay rate $\textit{$\kappa$}$. The real and complex parts of the eigenvalues of this matrix are used to calculate the excitation energy and excitation rate of the polariton modes in the two-mode stability analysis in the main text.

\section{Dicke model and the analytical expression for the excitation spectra}
To provide analytical insights into the mode softening of lower polariton at pump intensities lower than the critical value, we map the atom-cavity system onto the paradigmatic Dicke model. Specific details on the mapping can be found in Refs.~\cite{Skulte2021,Skulte_2024}. In the following, we simply highlight the important equations and the pertinent details are discussed in Ref.~\cite{Jara2024}. The open Dicke model is described by the Lindblad master equation \cite{Dimer2007},  
\begin{equation}
    \label{eq:dicke_master_eq}
    \partial_{t} \hat{\rho} = -i \left[ \frac{\hat{H}_{\mathrm{DM}}}{\hbar} , \rho \right] + \kappa \left(2\hat{a} \hat{\rho} \hat{a}^{\dagger} - \left\{ \hat{a}^{\dagger}\hat{a}, \hat{\rho}  \right\} \right),
\end{equation}
where the Hamiltonian reads
\begin{equation}
    \label{eq:dicke_model}
    \frac{\hat{H}_{\mathrm{DM}}}{\hbar} = \omega \hat{a}^{\dagger}\hat{a} + \omega_{0}\hat{J}_{z} + \frac{2\lambda}{\sqrt{N}} \left(\hat{a} + \hat{a}^{\dagger} \right) \hat{J}_{x} .
\end{equation}
The collective spin operators describing the matter sector approximated as two-level systems are $\hat{J}_\mu$, the light-matter coupling strength is $\lambda$, the two-level transition frequency is $\omega_0$, and the photon frequency is $\omega$. In terms of the original atom-cavity parameters, we have $\lambda = \sqrt{N_\mathrm{a}\omega_\mathrm{rec}\varepsilon |U_0|}/2$, where $\varepsilon$ is the unitless pump intensity, $\omega_0 = 2\omega_\mathrm{rec}$, and $\omega = -\delta_\mathrm{eff}$.
The critical light-matter coupling strength for the NP-SR phase transition is
\begin{equation}
    \label{eq:dicke_crit_point}
    \lambda_{\mathrm{c}} = \frac{1}{2} \sqrt{\frac{\omega_{0}}{\omega} \left( \kappa^{2} + \omega^{2} \right)  }.
\end{equation}
In the thermodynamic limit, we can use the Holstein-Primakoff transformation to obtain a linearised oscillator-like model (LOM) with a Hamiltonian given by
\begin{equation}
    \label{eq:dicke_lom_hamiltonian}
    \frac{\hat{H}_{\mathrm{LOM}}}{\hbar} = \omega \hat{a}^{\dagger}\hat{a} + \omega_{0} \hat{b}^{\dagger}\hat{b} + \lambda \left( \hat{a}^{\dagger} + \hat{a} \right) \left( \hat{b} + \hat{b}^{\dagger} \right),
\end{equation}
where $\hat{b}$ is the bosonic operator corresponding to the matter sector. This Hamiltonian is akin to that of a system with coupled linear oscillators. In particular, using a semi-classical treatment where we treat $\hat{a}$ and $\hat{b}$ as complex numbers, the equations of motion for $\hat{a}$ and $\hat{b}$ accounting for photon dissipation takes the form \cite{Jara2024}
\begin{equation}
\label{eq:lom_eom}
\frac{\partial^{2}}{\partial t}\textbf{x} + 
\begin{pmatrix}
2\kappa &  0 \\
0 		&  0 
\end{pmatrix} 
\frac{\partial}{\partial t}\textbf{x} + 
\begin{pmatrix}
\omega^{2} + \kappa^{2}	& 	2\lambda \sqrt{\omega \omega_{0}} \\
2\lambda \sqrt{\omega \omega_{0}}		& 	\omega_{0}^{2}
\end{pmatrix}
\textbf{x} = 0,
\end{equation} 
where $\textbf{x} = 
\begin{bmatrix}
x 	& 	y
\end{bmatrix}^{T}$ and
\begin{equation}
x = \frac{1}{\sqrt{2\omega}} (a + a^{*}), \quad p_{x} = i \sqrt{\frac{\omega}{2}}(a - a^{*}), \quad y = \frac{1}{\sqrt{2\omega_{0}}} (b + b^{*}), \quad p_{y} = i \sqrt{\frac{\omega_{0}}{2}}(b - b^{*}),
\end{equation}
with $a = \left< \hat{a} \right>$ and $b = \left< \hat{b} \right>$, $a, b \in \mathbb{C}$. In this mapping, obtaining the polariton modes of the atom-cavity system is equivalent to finding the normal modes of the LOM, in which each decoupled mode corresponds to a polariton mode \cite{Jara2024}. 

\begin{figure}
\centering
\includegraphics[width=0.7\linewidth]{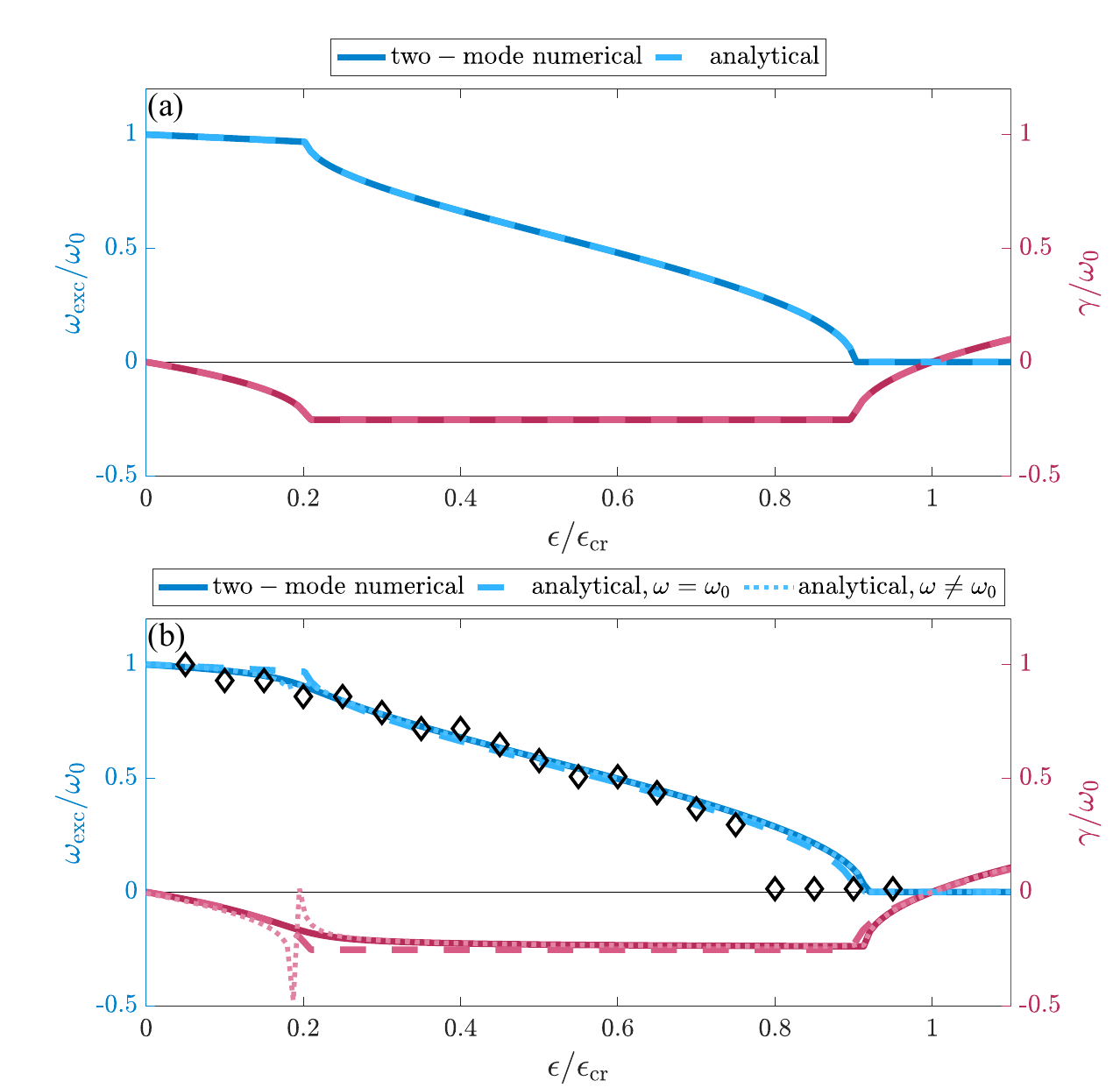}
\caption{Theoretical excitation rate and frequency for varying $\epsilon$ with (a) $|\delta_\mathrm{eff}| = 2\omega_\mathrm{rec}$ and (b) $\delta_\mathrm{eff} = -2\pi\times 7.5~\mathrm{kHz}$. The solid lines correspond to numerical diagonalisation of the stability matrix Eq.~\eqref{eq:stab_mat}. The broken lines correspond to the analytical predictions Eqs.~\eqref{eq:analytic} and \eqref{eq:analytic2}. The diamond markers in (b) denote the results from a full dynamical simulation of the multi-mode system discussed in the last section of the supplemental material.}
\label{sfig:1}
\end{figure}

We can derive the normal modes of the LOM by first substituting the Ansatz
\begin{equation}
\label{eq:normal_mode_ansatz}
\textbf{x} = \frac{1}{2}\exp(\Omega t)
\begin{pmatrix}
1 	& 	1 \\
1 	& 	-1
\end{pmatrix} 
\textbf{x}_{0} = \frac{1}{2}\exp \left(-\frac{\kappa t}{2}\right) \exp(\Gamma t)
\begin{pmatrix}
1 	& 	1 \\
1 	& 	-1
\end{pmatrix} 
\textbf{x}_{0},
\end{equation}
where $\textbf{x}_{0}$ is a constant vector, back to Eq.~\eqref{eq:lom_eom}. This leads to a nonlinear eigenvalue problem of the form,
\begin{equation}
\label{eq:eigval_problem}
\mathcal{M}(\Gamma)\textbf{x}_{0} = 0, \quad \mathcal{M}(\Gamma) \equiv
\begin{pmatrix}
\omega^{2}_{+} + \Gamma^{2} &  \omega^{2} - \frac{1}{4}\left( \Delta_{+}^{2} + \Delta_{-}^{2} \right) + \Gamma \kappa \\
\omega^{2} - \frac{1}{4}\left( \Delta_{+}^{2} + \Delta_{-}^{2} \right) + \Gamma \kappa & \omega_{-}^{2} + \Gamma^{2}
\end{pmatrix},
\end{equation}
where $\Delta_{\pm} = \omega \pm \omega_{0}$ and,
\begin{equation}
\omega_{\pm}^{2} = \frac{1}{4} \left( \Delta_{+}^{2} + \Delta_{-}^{2} \right) + \frac{\kappa^{2}}{4} \pm \lambda \sqrt{\Delta_{+}^{2} - \Delta_{-}^{2}}.
\end{equation} 
From Eq.~\eqref{eq:normal_mode_ansatz} and Eq.~\eqref{eq:eigval_problem}, it follows that the complex normal mode frequencies are
\begin{equation}
\label{eq:normal_mode_general}
\Omega = -\frac{\kappa}{2} + \Gamma_{0}(\omega, \omega_{0}),
\end{equation} 
where $\Gamma_{0}(\omega, \omega_{0})$ are the complex roots of the characteristic polynomial of $\mathcal{M}(\Gamma)$, defined as
\begin{subequations}
\label{eq:polariton_characteristic_eq}
\begin{equation}
F(\Gamma) \equiv \det (\mathcal{M}(\Gamma)),
\end{equation}
\begin{equation}
F(\Gamma)= \left[ \Gamma^{2} + \frac{1}{2}\left( \omega^{2} + \omega_{0}^{2} \right) -\frac{\kappa^{2}}{4} \right]^{2} - \Gamma \kappa \left(\omega^{2} - \omega_{0}^{2} \right) - \lambda'^{2}\omega_{0}^{2}\left( \kappa^{2} + \omega^{2} \right) + \frac{1}{2} \left( \omega^{2} + \omega_{0}^{2} \right)\kappa^{2} - \frac{1}{4}\left( \omega^{2} - \omega_{0}^{2} \right)^{2}.
\end{equation}
\end{subequations}
An analytical expression can be obtained in the resonant regime $\omega=\omega_0$ and, focusing on the lower polariton, the corresponding complex lower polariton frequency is given by
\begin{equation}\label{eq:analytic}
\Omega_\mathrm{LP} = -\frac{\kappa}{2} + \Omega_0,
\end{equation} 
where
\begin{equation}\label{eq:bare}
\Omega_0 = \sqrt{\frac{\kappa^2}{4} - \omega^2 + \omega\sqrt{\lambda'^2(\kappa^2+\omega^2)-\kappa^2}}
\end{equation}
and $\lambda' = \lambda/\lambda_\mathrm{c}$. This expression is consistent with that from an eigenvalue analysis of the Lindbladian in Ref.~\cite{Dimer2007}.
Note that the structure of the normal mode solution being $\mathbf{x} \propto \exp(\Omega_\mathrm{LP} t) \mathbf{x_0} $ means that the \textbf{\textit{excitation frequency}} is $\omega_\mathrm{exc}=\mathrm{Im}[\Omega_\mathrm{LP}]$ and the \textbf{\textit{excitation rate}} is $\gamma=\mathrm{Re}[\Omega_\mathrm{LP}]$, which is opposite to the two-mode stability analysis described above. We point out this difference is just a matter of convention and that these two approaches can be made consistent by simply multiplying $\Omega_\mathrm{LP}$ with $i$. 

In SFig.~\ref{sfig:1}(a), we present a good agreement between the numerical diagonalisation of the two-mode stability matrix Eq.~\eqref{eq:stab_mat} and the analytical expression Eq.~\eqref{eq:analytic} for the excitation frequency and rate in the resonant regime. From Eqs.~\eqref{eq:analytic} and \eqref{eq:bare}, we can infer that $\gamma$ becomes zero at the critical point $\lambda'=1$ as expected. More importantly, we can see that the excitation frequency is solely dependent on $\Omega_0$ in the resonant regime since $\omega_\mathrm{exc} = \mathrm{Im}[\Omega_\mathrm{LP}] =\mathrm{Im}[\Omega_0] $. In the absence of dissipation the ``bare" polariton frequency is $\Omega_\mathrm{LP}=\Omega_0 = \omega\sqrt{\lambda' -1}$ and we find that both the excitation energy and rate become zero at the critical point. We can then interpret Eq.~\eqref{eq:analytic} as the dissipation not only modifying the bare polariton frequency $\Omega_0$ but also, more importantly for the existence of the polaritonic zero-energy excitation mode, effectively decreasing the excitation rate by shifting the excitation rate curve by half of the dissipation rate $\kappa/2$. This leads to the excitation rate $\gamma$ no longer coinciding with the excitation frequency $\omega_\mathrm{exc}$ at $\lambda = \lambda_\mathrm{cr}$. As the zero-crossing of $\gamma$ gets pushed to higher values of $\lambda$ owing to the dissipation-induced negative shift, the excitation frequency  $\omega_\mathrm{exc}$ becomes zero for $\lambda < \lambda_\mathrm{c}$. The apparent mode softening of the lower polariton below the critical point can thus be attributed to dissipation.

\begin{figure}
\centering
\includegraphics[width=0.7\linewidth]{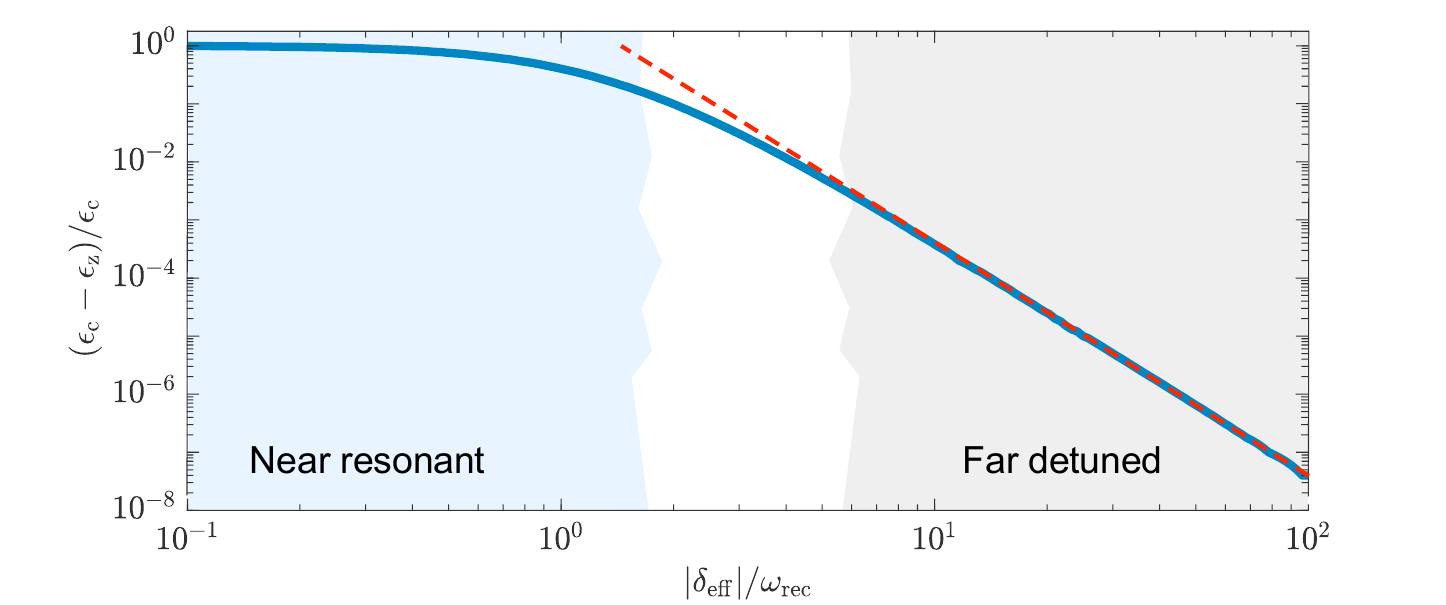}
\caption{$\Delta\epsilon_\mathrm{z} \equiv 1-\frac{\epsilon_\mathrm{z}}{\epsilon_\mathrm{c}}$ is plotted versus $|\delta_\mathrm{eff}| / \omega_\mathrm{rec}$ across the near resonant and the far detuned regime. The red dashed line shows an excellent fit with a $[\delta_\mathrm{eff} / \omega_\mathrm{rec}]^{-4}$ power law.}
\label{sfig:2}
\end{figure}

While we can obtain an analytic expression for the polariton frequencies in the resonant regime, finding the roots of $F(\Gamma)$ in the general case of $\omega \neq \omega_{0}$ becomes intractable due to the presence of the $\Gamma \kappa(\omega^{2}-\omega_{0}^{2})$ cross-term. To obtain a closed-form approximation for the $\Gamma_{0}$ corresponding to the lower polariton in the small detuning case, we use the first iteration of the Newton-Rhapson method, where we assume that
\begin{equation}
\label{eq:root_approx}
\Gamma_{0, \mathrm{LP}}(\omega \neq \omega_{0}) \approx \Omega_{0} - \frac{F\left(\Omega_{0}\right)}{\partial_{\Gamma}F(\Gamma) |_{\Gamma = \Omega_{0}} }, 
\end{equation}
with
\begin{equation}
F(\Omega_{0}) = \left( \omega^{2} - \omega_{0}^{2} \right)\left[\lambda'^{2}\left( \kappa^{2} + \omega^{2} \right) - \frac{\kappa^{2}}{2} - \omega \sqrt{\lambda'^{2}(\kappa^{2} + \omega^{2}) - \kappa^{2}} - \Omega_{0} \kappa   \right],
\end{equation}
and
\begin{equation}
\left. \frac{\partial F(\Gamma)}{\partial \Gamma}\right|_{\Gamma = \Omega_{0}} = 4\omega \Omega_{0}\sqrt{\lambda'^{2}(\kappa^{2} + \omega^{2}) - \kappa^{2}} - \left(\omega^{2} - \omega_{0}^{2} \right)\left(\kappa + 2\Omega_{0} \right).
\end{equation}
This assumption is justified since as the detuning between $\omega$ and $\omega_{0}$ goes to zero, $\Delta_{-} \rightarrow 0$, $\Gamma_{0} \rightarrow \Omega_{0}$. Thus, Eq.~\eqref{eq:root_approx} should provide a good approximation for $\Gamma_{0}$ so long as the detuning is sufficiently small. With this, the lower polariton frequency for small detuning finally reads,
\begin{equation}\label{eq:analytic2}
\Omega_\mathrm{LP} = -\frac{\kappa}{2} + \Omega_0 - (\omega^2-\omega_0^2) \Omega'
\end{equation}
where the correction due to the detuning between $\omega$ and $\omega_0$ is
\begin{equation}
\Omega' = \frac{\lambda'^2(\kappa^2+\omega^2)-\kappa^2/2 - \omega\sqrt{\lambda'^2(\kappa^2+\omega^2)-\kappa^2}-\Omega_0 \kappa}{4\omega \Omega_0\sqrt{\lambda'^2(\kappa^2+\omega^2)-\kappa^2} - (\omega^2-\omega_0^2)(\kappa+2\Omega_0)}.
\end{equation}
Note that by virtue of Eq.~\eqref{eq:normal_mode_general}, the general structure of $\Omega_\mathrm{LP}$ remains the same as in the resonant regime, wherein the dissipation introduces the $-\kappa/2$ shift in the excitation rate. This is further corroborated by the good agreement between $\omega=\omega_0$ and $\omega\neq \omega_0$ in the SFig.~\ref{sfig:1}(b), which supports the interpretation that the existence of the polaritonic zero-energy excitation is due to dissipation. We also point out that the diamond markers in SFig.~\ref{sfig:1}(b), which correspond to the numerical results from a full dynamical simulation of the longitudinal probing protocol using a multi-mode atom-cavity model (this will be discussed in the last section of the supplemental material), are in good agreement with the two-mode stability matrix simulation and the analytical results as given by Eq.~\eqref{eq:analytic2}, especially for small pump intensities. This further underpins the underlying dissipation-induced mechanism for the pre-critical point mode softening of the lower polariton. The deviations for larger $\epsilon$ suggest that the higher-momentum modes neglected in the two-mode and analytical descriptions become relevant in this regime. Nevertheless, the various theoretical approaches all agree that there is a polaritonic zero-energy excitation in the system.

\section{Regimes of R-LIRI}
The width of the pump strength interval $[\epsilon_\mathrm{z},\epsilon_\mathrm{c}]$, across which the excitation energy of the lowest polariton mode becomes zero, scaled to the critical pump strength $\textit{$\epsilon$}_\mathrm{c}$, determines the degree of experimental observability of the zero-energy mode behavior. This quantity, given by $\Delta\epsilon_\mathrm{z} \equiv 1-\frac{\epsilon_\mathrm{z}}{\epsilon_\mathrm{c}}$ approaches zero in the far detuned regime, where $|\delta_\mathrm{eff}| \gg \omega_\mathrm{rec}$. In this regime, the zero-excitation-energy mode character is thus practically undetectable. All experiments besides ours operate deeply within this regime due to the use of values for $\kappa$ exceeding the recoil frequency by more than two orders of magnitude. In contrast, in the near resonant regime, when $2 \omega_\mathrm{rec} \gtrsim |\delta_\mathrm{eff}|$, $\Delta\epsilon_\mathrm{z}$ approaches unity and the zero-energy character of the mode becomes detectable. In SFig.~\ref{sfig:2}, $\Delta\epsilon_\mathrm{z} \equiv 1-\frac{\epsilon_\mathrm{z}}{\epsilon_\mathrm{c}}$ is plotted versus $|\delta_\mathrm{eff}| / \omega_\mathrm{rec}$ across the near resonant and the far detuned regime. The red dashed line shows an excellent fit with a $[\delta_\mathrm{eff} / \omega_\mathrm{rec}]^{-4}$ power law.

\section{Experimental details}
The experimental setup consists of a Bose-Einstein condensate of $N_\mathrm{a}=40\cdot10^3$ \textsuperscript{87}Rb atoms, magnetically confined in a magnetic trap with harmonic trap frequencies of $(\textit{$\omega$}_\mathrm{x},\textit{$\omega$}_y,\textit{$\omega$}_z)=2\textit{$\pi$}\cdot(83.0,72.2,23.7)$ \si{Hz}. These atoms are prepared in the $\ket{5^2S\textsubscript{1/2},F=2,m_f=2}$ hyperfine substate and coupled to a single mode of a high-finesse recoil-resolved optical cavity with a field decay rate of $\textit{$\kappa$}=2\textit{$\pi$}\cdot3.60$ \si{\kilo Hz}, which is on the same scale as the recoil frequency $\textit{$\omega$}_\mathrm{rec}=2\textit{$\pi$}\cdot3.55$ \si{\kilo Hz} at the chosen pump wavelength $\textit{$\lambda$}_\mathrm{p}$. The transverse pump, which is used to control the coupling strength between the atoms and the cavity mode, is set to a wavelength of $\textit{$\lambda$}_\mathrm{p}=803.26$ \si{\nano m}, far-red detuned with respect to the relevant atomic transitions at $794.98$ and $780.24$ \si{\nano m}, thus we operate in the dispersive regime. For a light shift per photon of $U_\mathrm{0}=-2\textit{$\pi$}\cdot0.34$ \si{Hz}, the total dispersive shift for the coupling with the $\textit{$\sigma$}^-$ polarisation is equal to $\textit{$\delta$}_-=\frac{N_\mathrm{a}U_\mathrm{0}}{2}$. Thus, we obtain the effective detuning $\textit{$\delta$}_\mathrm{eff}=\textit{$\delta$}_\mathrm{c}-\textit{$\delta$}_-$, containing the dispersive shift.

\section{Cavity field detection}
In our experiment, we use two methods to detect the light leaking out of the cavity. On one side, we detect the light transmitted by the high-reflecting mirror with a single-photon counting module (SPCM), which allows us to measure the intensity of the transmitted intracavity field and longitudinal probe signal. On the other side of the cavity, we detect the cavity leakage light using a balanced heterodyne setup, which employs a split-off branch of the transverse pump beam as a local reference. The beating signal generated by these two beams allows us to determine the time-resolved intracavity photon number $N_\mathrm{p}(t)$ and the phase difference between the pump beam and the intracavity light field.

\section{Experimental protocol and data evaluations}
\textbf{Protocol:}\\
To experimentally determine the excitation energy at a fraction of the critical pump strength, $\textit{$\epsilon$}_\mathrm{p}<\textit{$\epsilon$}_\mathrm{c}$, we first determine the value of critical pump strength for a fixed detuning $\delta_{\mathrm{eff}}$. Thus, we fix the detuning $\delta_{\mathrm{eff}}$ for an entire excitation spectrum, load the BEC into the cavity, and linearly ramp the transverse pump strength $\textit{$\epsilon$}_\mathrm{p}$ over the critical threshold with a constant gradient, without any involvement of the longitudinal probe. This measurement is repeated between measurements of the excitation energy to compensate for drifts in $\textit{$\epsilon$}_\mathrm{c}$ over time. This is done by calculating the compensated transverse pump strength $\textit{$\epsilon$}_\mathrm{p}$ as a fraction of the critical pump strength $\textit{$\epsilon$}_\mathrm{c}$ from the mean of the two closest critical pump strength measurements before and after the fixed transverse pump strength is probed. The standard deviation between these two data points and the error on the individual critical pump strength measurements result in an uncertainty when determining the exact fraction of the critical pump strength $\textit{$\epsilon$}_\mathrm{p}/\textit{$\epsilon$}_\mathrm{c}$, which is the cause for horizontal error bars in the excitation spectra plots. Following this, to measure the excitation energy of the polariton mode, we load the BEC into the cavity and linearly increase the transverse pump to a fraction of the critical pump strength using a constant gradient and hold the pump strength at this value for $20$ \si{\milli s}. After the ramp, we activate the longitudinal probe, linearly sweep the detuning between the longitudinal probe and the transverse pump from $\textit{$\delta$}_\mathrm{lt}/2\textit{$\pi$}=-10$ to $+1$ \si{\kilo Hz}, and measure the photon number $N_\mathrm{p}$ in the cavity by detecting the light leaking out of the cavity using a balanced heterodyne detector.\\ 

\begin{figure}
\centering
\includegraphics[width=0.5\linewidth]{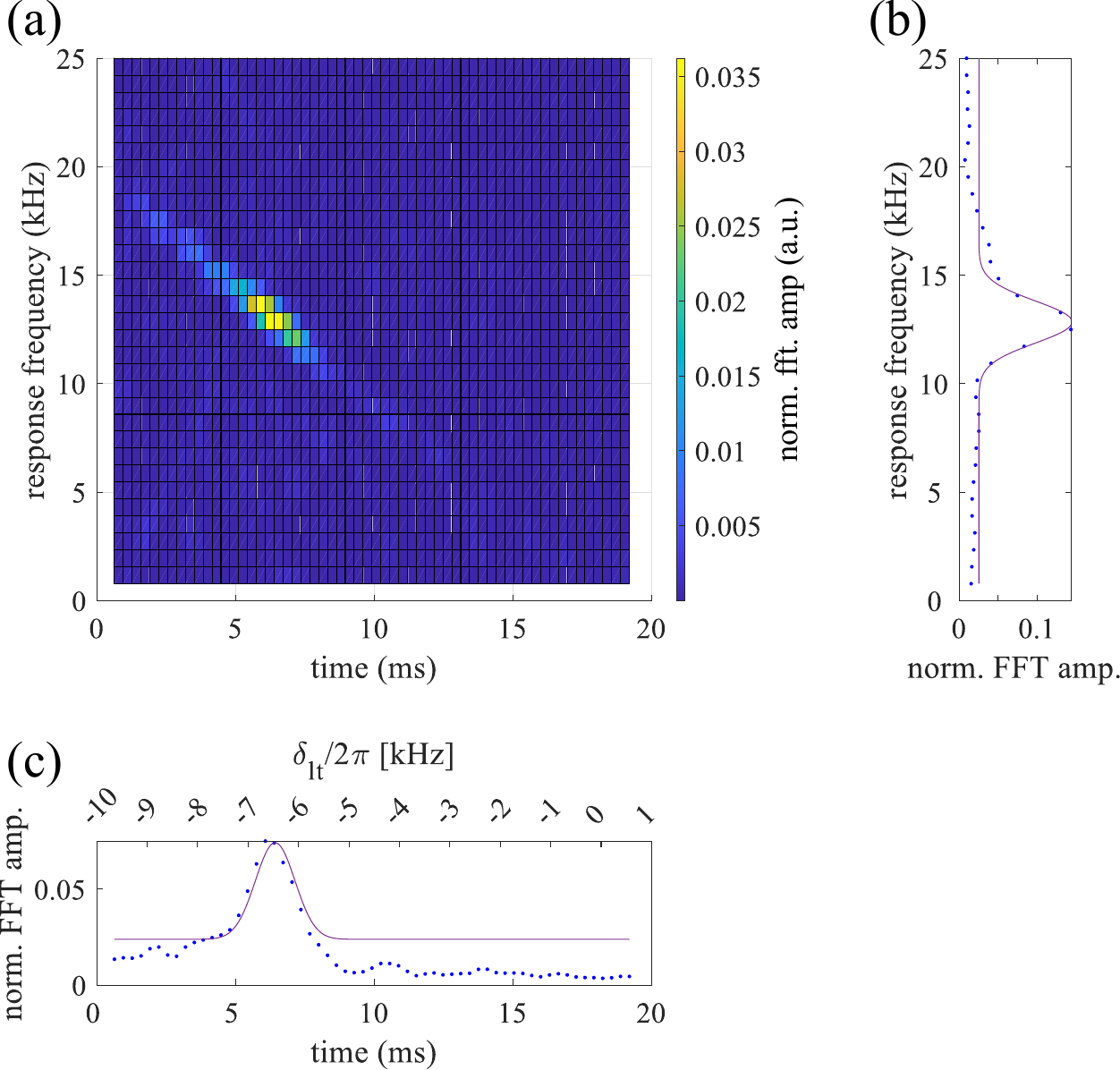}
\caption{(a) Single-sided spectrogram obtained from the time-resolved Fourier transform of the photon number $N_\mathrm{p}$ measured by the heterodyne detector for an effective detuning of $\textit{$\delta$}_\mathrm{eff}/2\textit{$\pi$}=-15$ \si{\kilo Hz} and a transverse pump strength of $\textit{$\epsilon$}_\mathrm{p}/\textit{$\epsilon$}_\mathrm{c}=0.2$. (b) The sum of the normalised Fourier amplitudes within a given frequency range for all time bins is plotted against the response frequency. (c) The sum of the normalised Fourier amplitudes for all frequencies within one time bin plotted against the measurement time and detuning $\textit{$\delta$}_\mathrm{lt}$. A Gaussian profile is fitted to (b) and (c) to separately determine the excitation energy from each method. The result from (c) is used to determine the excitation energy in the results shown here due to its higher precision, while the result from (b) is used as a reference value for post-selection.}
\label{sfig:3}
\end{figure}

\textbf{Data evaluation:}\\
We apply a digital bandpass filter with a pass range between $500$ \si{Hz} and $100$ \si{\kilo Hz} to the detected photon number to remove both high-frequency noise and the offset added to the signal by the longitudinal probe light reflected by the cavity coupling mirror. After this, we apply a time-resolved Fourier transform to the filtered photon number, binning the data into 59 intervals and analysing the Fourier amplitude for the given frequency components during the sweep of $\textit{$\delta$}_\mathrm{lt}$, see SFig.~\ref{sfig:3}(a). At resonance, we would expect a beating frequency of $2\textit{$\omega$}_\mathrm{exc}$, as the longitudinal probe and the transverse pump beat, causing a density modulation at the position of the atoms of this frequency, which is then transferred to the light field. In addition to this, the signal, which has the frequency of the transverse pump, beats with the longitudinal probe on the heterodyne detector. These two modulations cause the photon number to oscillate at $2\textit{$\omega$}_\mathrm{exc}$ when the resonance condition is reached, and light is scattered from the density lattice into the cavity. We obtain the excitation energy $\textit{$\omega$}_\mathrm{exc}$ of the polariton mode from the detuning $\textit{$\delta$}_\mathrm{lt}$ that corresponds to the time during the scan at which the most intracavity photon is detected, see the bright area on the spectogram in SFig.~\ref{sfig:3}(a). This measurement is repeated between $10$ and $20$ times for a fixed transverse pump strength. Averaging over the results of the excitation energies from these measurements, we determine $\textit{$\omega$}_\mathrm{exc}$ as presented in Fig.~3 of the main text and SFig.~\ref{sfig:6}. The vertical error bars in the individual data points accounting for the standard deviation of the fitted excitation energies around the mean as well as the respective uncertainties from the fit function. The process is repeated for different fractions of $\textit{$\epsilon$}_\mathrm{p}/\textit{$\epsilon$}_\mathrm{c}$ to obtain a full excitation spectrum.\\ 

\textbf{Post-selection process:}\\
To prevent data that does not accurately determine the excitation energy of a polariton mode from obscuring the results of our measurements, we employ post-selection criteria to filter these data out. The main cause of erroneous data is measurements that got excited into the SR phase, especially when the chosen $\textit{$\epsilon$}_\mathrm{p}$ is close to $\textit{$\epsilon$}_\mathrm{c}$, where atom number fluctuations can easily cause the system to enter self-organisation. Such data cannot be used to extract the excitation energy. In self-organisation, light is scattered into the cavity throughout the measurement, producing a range of Fourier spectrum. The erroneous data is filtered out by comparing the excitation energy calculated by summing over the normalised Fourier amplitudes in frequency and time respectively, as is shown in SFigs.~\ref{sfig:3}(b-c). If the excitation energies obtained from these Gaussian fits differ significantly, with our cutoff being $1$ \si{\kilo Hz}, we cannot verify that what we measured is an excitation of the polariton mode, and the result of this individual data point is removed from further analysis. Another employed post-selection method, is to make a cut-off in the signal-to-noise ratio of the normalised Fourier amplitude. Further criteria include the goodness of fit for the Gaussian fit function and the error on the excitation energy from the fit function to ensure that the data corresponds to the accurate fit of a single excitation peak. 

\section{Intensity compensation scheme}

\begin{figure}[H]
	\centering
	\includegraphics[width=0.4\linewidth]{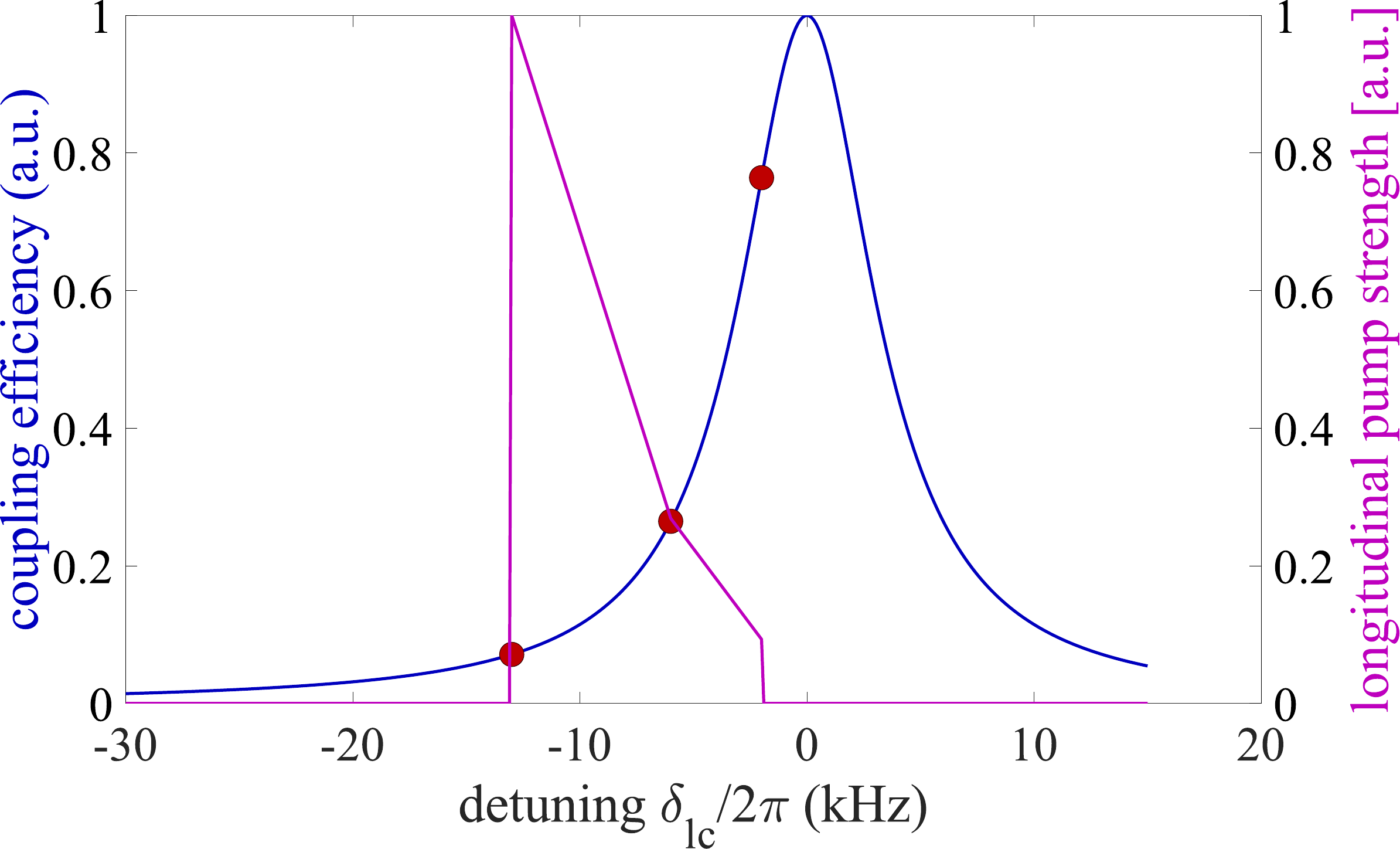}
	\caption{Intensity compensation scheme applied to the longitudinal probe intensity to address the frequency-dependent coupling efficiency of the cavity for an effective detuning of $\textit{$\delta$}_\mathrm{eff}/2\textit{$\pi$}=-3$ \si{\kilo Hz}. \textbf{Blue:} modelled frequency-dependent coupling efficiency of the cavity for the $TEM_{00}$ mode plotted against the detuning $\textit{$\delta$}_{lc}$ between the longitudinal probe and the effective cavity resonance frequency, \textbf{red:} marked points in the scan range between which the longitudinal probe intensity is linearly decreased, and \textbf{pink:} modelled longitudinal probe intensity after applying the compensation protocol.}
	\label{sfig:4}
\end{figure}

In our experiment, we use an on-axis longitudinal probe to measure the excitation energy of the polariton modes. As we sweep the frequency of the longitudinal probe with respect to the transverse pump frequency, the coupling efficiency of the longitudinal probe with respect to the cavity changes, see the blue curve in SFig.~\ref{sfig:4}. The goal of this scheme is to maintain the same coupling efficiency throughout the frequency sweep protocol by modifying the longitudinal probe strength to compensate for the differences in coupling efficiency. An example of the compensated longitudinal probe power can be seen in the pink curve SFig.~\ref{sfig:4}.\\

The protocol works as follows: we calculate the inverse of the coupling efficiency of three points in the scan, here at $\textit{$\delta$}_{lc}/2\textit{$\pi$}=[-13,-6,-2]$ \si{\kilo Hz} for a fixed effective transverse pump detuning, here as an example, we use $\textit{$\delta$}_\mathrm{eff}/2\textit{$\pi$}=-3$ \si{\kilo Hz}, which are $[0.0712,0.2646,0.7634]$ and the inverse $[14.04,3.78,1.31]$. These inverse values of the coupling efficiencies are normalised by dividing them by the smallest value, leading to the following compensation factors: $[10.72,2.89,1.00]$. We then return to the three detunings $\textit{$\delta$}_\mathrm{lt}$ marked as red dots in SFig.~\ref{sfig:4}, multiply the longitudinal probe strength set on the experimental control by the value corresponding to the frequency of those points, and apply a linear gradient of the longitudinal probe strength between neighbouring values, results of the compensated probe power is shown in SFig.~\ref{sfig:4}. A linear ramp is chosen for its ease of implementation in the experimental control and its relative effectiveness.\\

\begin{figure}[h]
    \centering
    \includegraphics[width=0.4\linewidth]{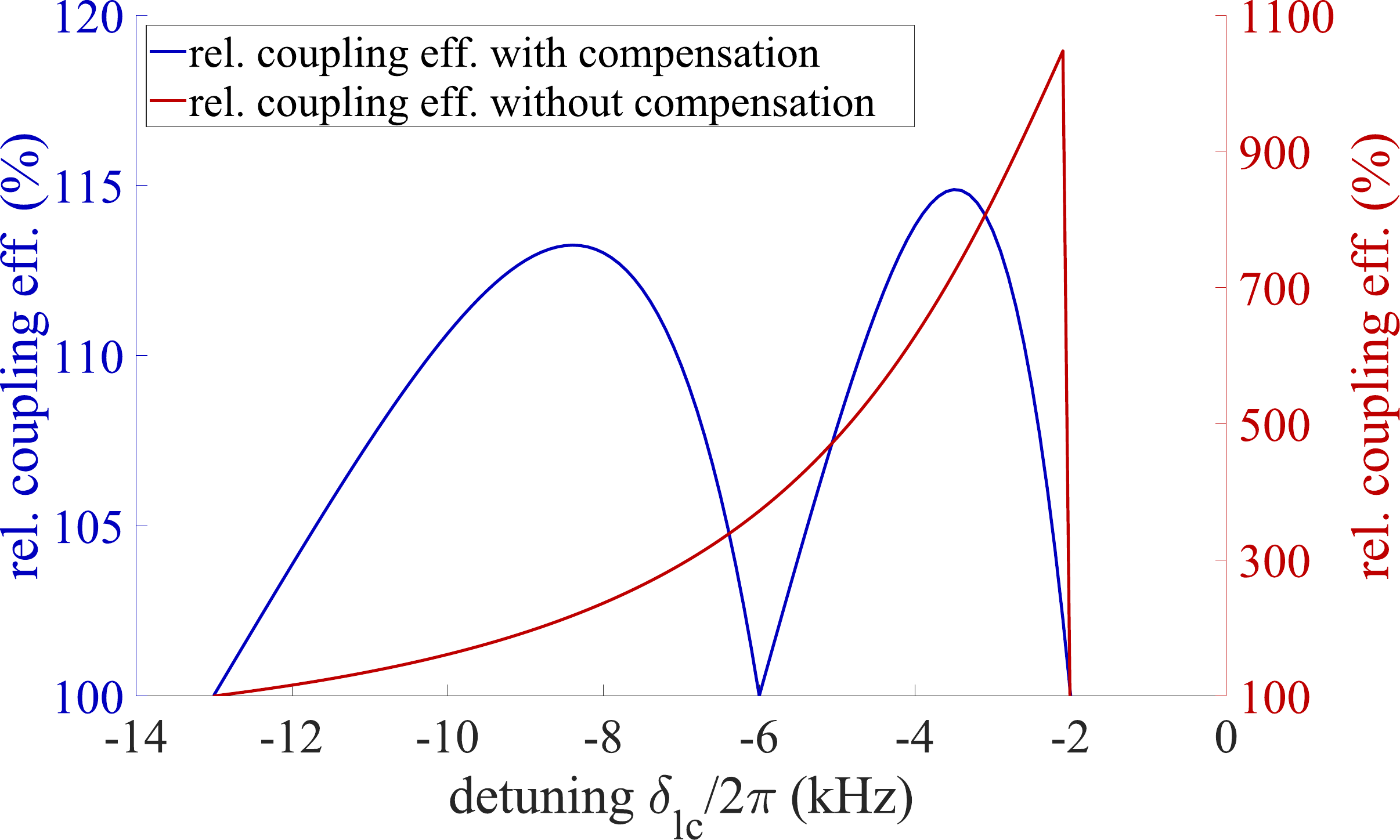}
    \caption{Modelled coupling efficiency with (blue) and without (red) application of the intensity compensation protocol, as presented in SFig.~\ref{sfig:4}, for an effective detuning of $\textit{$\delta$}_\mathrm{eff}/2\textit{$\pi$}=-3$ \si{\kilo Hz}. The coupling efficiency is relative to the value at $\textit{$\delta$}_\mathrm{lt}/2\textit{$\pi$}=-13$ \si{\kilo Hz}.}
    \label{sfig:5}
\end{figure}

The relative coupling efficiency calculated with and without the compensated probe power are shown in SFig.~\ref{sfig:5}. Without the compensation scheme, the differences in coupling efficiency throughout the scan exceeds almost 10 times close to the cavity resonance, see red curve in SFig.~\ref{sfig:5}. However, with the compensating probe power, three points in the scan at $\textit{$\delta$}_{lc}/2\textit{$\pi$}=[-13,-6,-2]$ \si{\kilo Hz} have their relative coupling efficiency at 100$\%$ and their neighbouring scan frequencies have only maximum deviation of around $14.9\%$. An ideal scenario is to have the relative coupling efficiency at 100$\%$ throughout the frequency scan. One can improve the compensating scheme to maintain the exact same coupling efficiency throughout the frequency sweep by increasing the number of compensated points in $\textit{$\delta$}_{lc}/2\textit{$\pi$}$ or applying a different interpolating function than a linear gradient.  A linear ramp is chosen in our case  for its ease of implementation in the experimental control and its relative effectiveness, as can be seen in SFig.~\ref{sfig:5}.


\section{Additional excitation spectra}

\begin{figure}[H]
\centering
\includegraphics[width=0.6\linewidth]{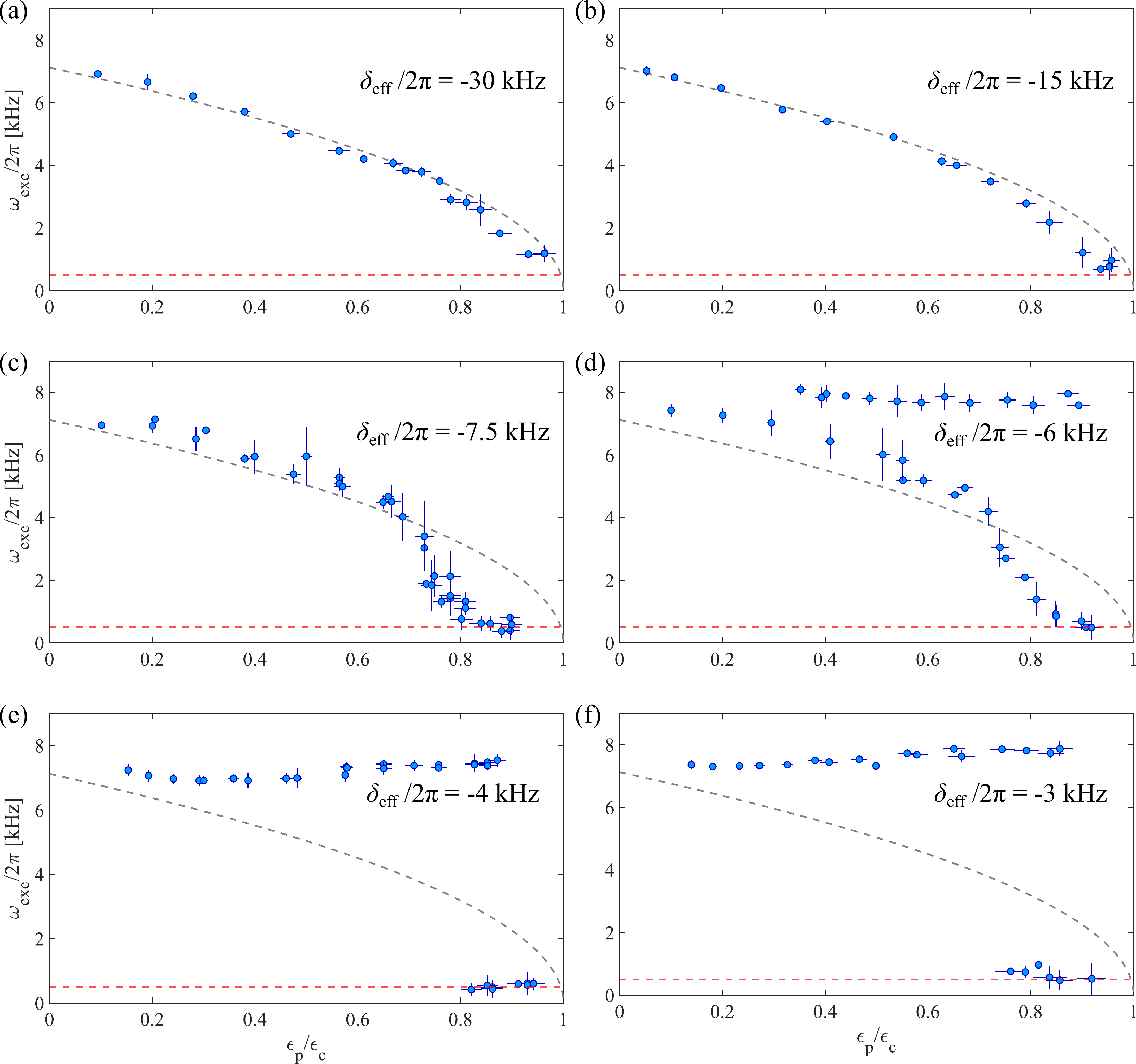}
\caption{Excitation spectra measured for effective detunings of (a) $\textit{$\delta$}_\mathrm{eff}/2\textit{$\pi$}=-30$ \si{\kilo Hz}, (b) $\textit{$\delta$}_\mathrm{eff}/2\textit{$\pi$}=-15$ \si{\kilo Hz}, (c) $\textit{$\delta$}_\mathrm{eff}/2\textit{$\pi$}=-7.5$ \si{\kilo Hz}, (d) $\textit{$\delta$}_\mathrm{eff}/2\textit{$\pi$}=-6$ \si{\kilo Hz}, (e) $\textit{$\delta$}_\mathrm{eff}/2\textit{$\pi$}=-4$ \si{\kilo Hz}, and (f) $\textit{$\delta$}_\mathrm{eff}/2\textit{$\pi$}=-3$ \si{\kilo Hz} as the transverse pump strength $\textit{$\epsilon$}_\mathrm{p}$ approaches the self-organisation threshold $\textit{$\epsilon$}_\mathrm{c}$, showing the progression of the excitation spectrum from the far-detuned to the near-detuned regime. The lowest frequency that can accurately be resolved with our experimental protocol, $500$ \si{Hz}, is marked by a red dashed line, and the square-root behaviour predicted for the far-detuned regime is shown in grey.}
\label{sfig:6}
\end{figure}

In addition to the results shown in the main manuscript, we recorded excitation spectra at effective detunings of $\textit{$\delta$}_\mathrm{eff}/2\textit{$\pi$}=-15$ and $-4$ \si{\kilo Hz}, which are included alongside the spectra discussed in the main text in SFig.~\ref{sfig:6}(b) and (e) for completeness. For an effective detuning of  $\textit{$\delta$}_\mathrm{eff}/2\textit{$\pi$}=-15$ \si{\kilo Hz}, we observe that the deviation from the square-root behaviour predicted for a cavity with significantly larger $\textit{$\kappa$}$ are more pronounced, the excitation spectrum at this detuning shows an  intermediate stage between the excitation spectra at $\textit{$\delta$}_\mathrm{eff}/2\textit{$\pi$}=-30$ and $\textit{$\delta$}_\mathrm{eff}/2\textit{$\pi$}=-7.5$ \si{\kilo Hz}. The excitation spectrum at $\textit{$\delta$}_\mathrm{eff}/2\textit{$\pi$}=-4$ \si{\kilo Hz} shows behaviour similar to the previously discussed results at $\textit{$\delta$}_\mathrm{eff}/2\textit{$\pi$}=-3$ \si{\kilo Hz}, as both contain an upper polariton mode present throughout the entire spectrum and a lower mode with an excitation frequency close to $0$ \si{\kilo Hz}.

\section{Dynamic multi-mode simulations}

\begin{figure}[h]
\centering
\includegraphics[width=0.85\linewidth]{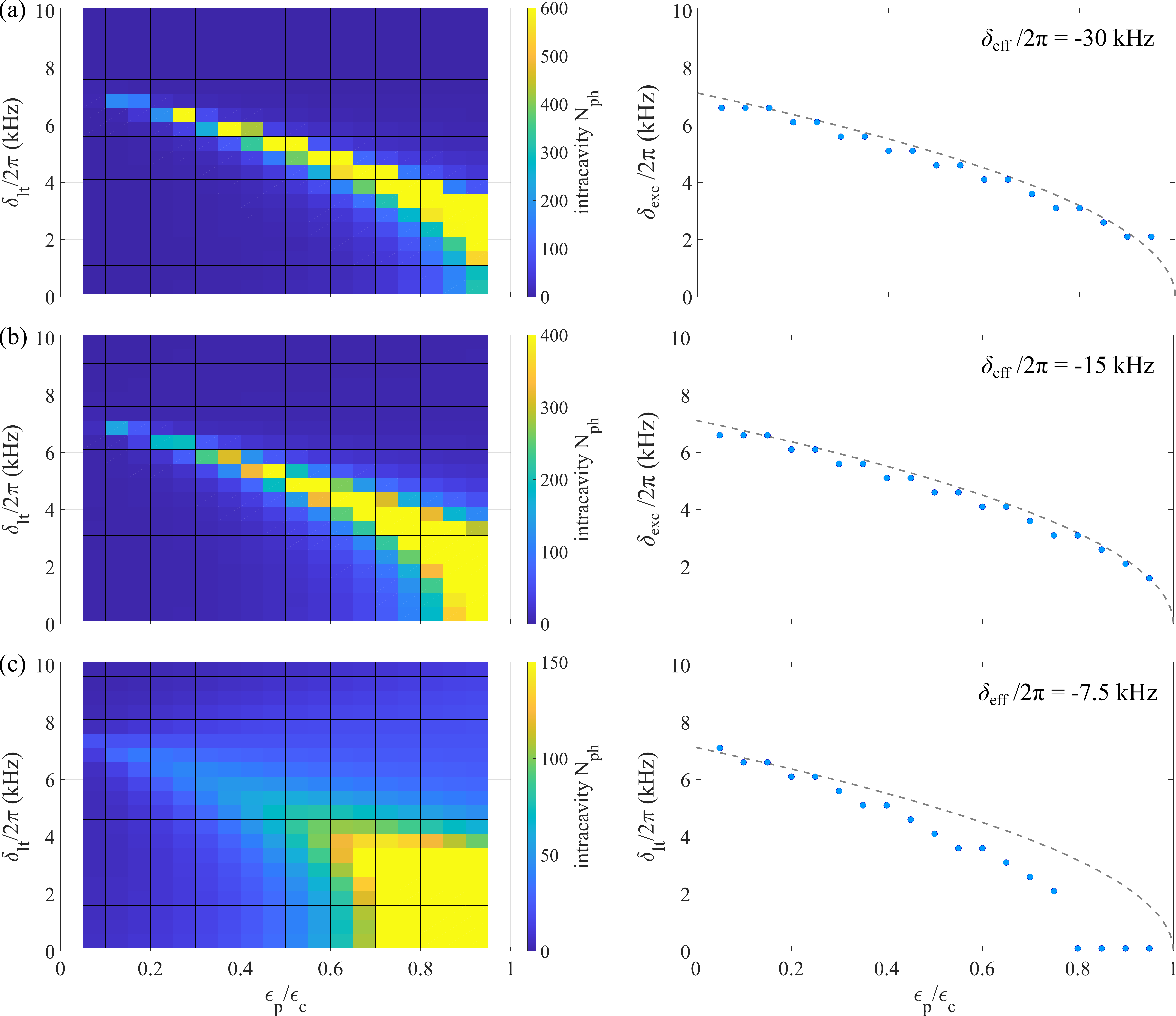}
\caption{Intracavity photon number resulting from the dynamic multimode simulations when varying the transverse pump strength $\textit{$\epsilon$}_\mathrm{p}/\textit{$\epsilon$}_\mathrm{c}$ and the detuning between the longitudinal probe and the transverse pump $\textit{$\delta$}_\mathrm{lt}$ for effective detunings above $2\textit{$\omega$}_\mathrm{rec}$. (a) $\textit{$\delta$}_\mathrm{eff}/2\textit{$\pi$}=-30$ \si{\kilo Hz}, (b) $\textit{$\delta$}_\mathrm{eff}/2\textit{$\pi$}=-15$ \si{\kilo Hz},  and (c) $\textit{$\delta$}_\mathrm{eff}/2\textit{$\pi$}=-7.5$ \si{\kilo Hz}. \textbf{Left:} total intracavity photon number for each $\textit{$\delta$}_\mathrm{lt}$, and \textbf{right:} $\textit{$\delta$}_\mathrm{lt}$ at which the most longitudinal probe photons are scattered into the cavity for a fixed transverse pump strength $\textit{$\epsilon$}_\mathrm{p}/\textit{$\epsilon$}_\mathrm{c}$.}
\label{fig:7}
\end{figure}

\begin{figure}[h]
\centering
\includegraphics[width=0.85\linewidth]{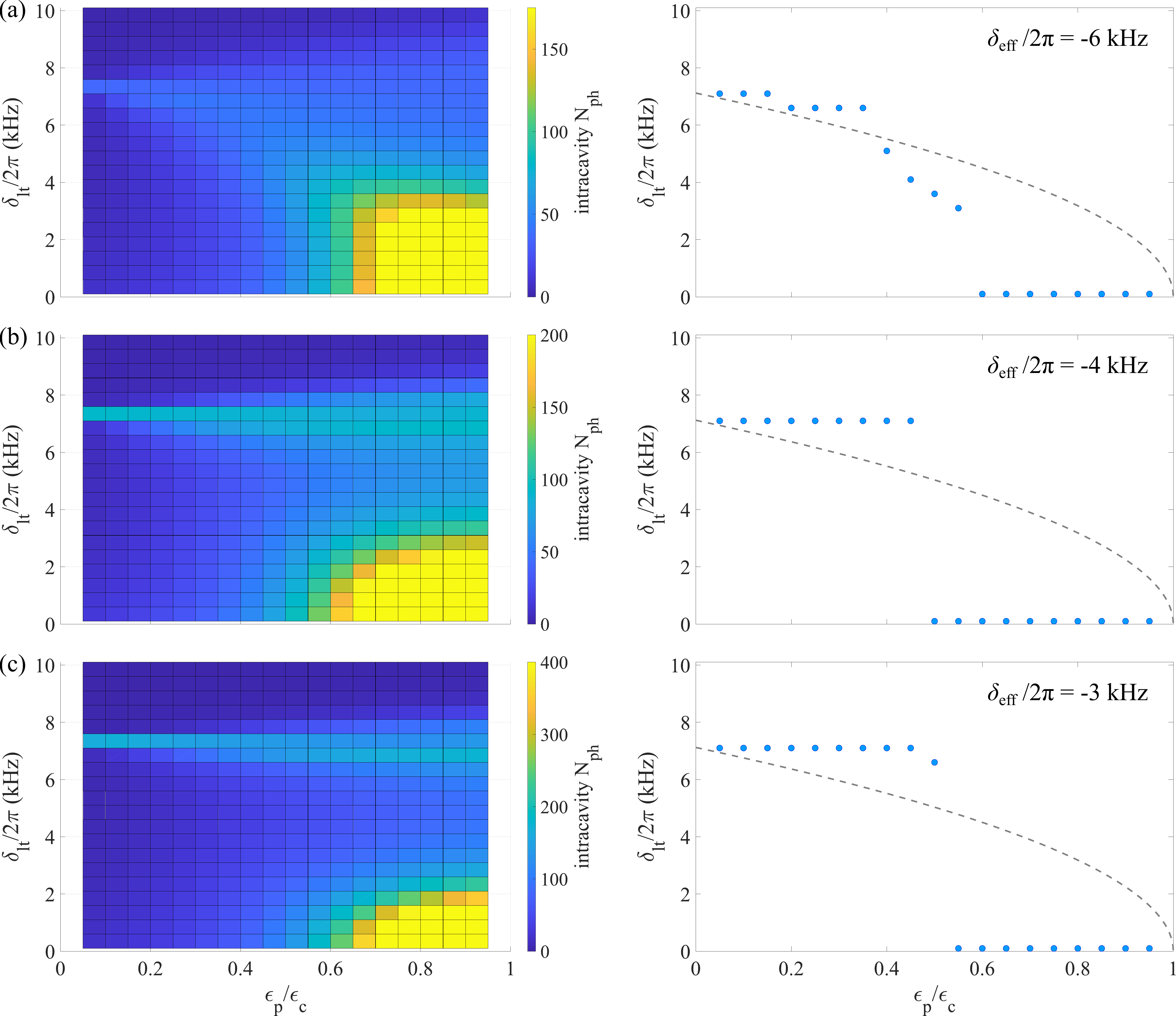}
\caption{Intracavity photon number resulting from the dynamic multimode simulations when varying the transverse pump strength $\textit{$\epsilon$}_\mathrm{p}/\textit{$\epsilon$}_\mathrm{c}$ and the detuning between the longitudinal probe and the transverse pump $\textit{$\delta$}_\mathrm{lt}$ for effective detunings below $2\textit{$\omega$}_\mathrm{rec}$. (a) $\textit{$\delta$}_\mathrm{eff}/2\textit{$\pi$}=-6$ \si{\kilo Hz}, (b) $\textit{$\delta$}_\mathrm{eff}/2\textit{$\pi$}=-4$ \si{\kilo Hz}, and (c) $\textit{$\delta$}_\mathrm{eff}/2\textit{$\pi$}=-3$ \si{\kilo Hz}. \textbf{Left:} total intracavity photon number for each $\textit{$\delta$}_\mathrm{lt}$, and \textbf{right:} $\textit{$\delta$}_\mathrm{lt}$ at which the most longitudinal probe photons are scattered into the cavity for a fixed transverse pump strength $\textit{$\epsilon$}_\mathrm{p}/\textit{$\epsilon$}_\mathrm{c}$.}
\label{fig:8}
\end{figure}

To simulate the exact conditions of our atom-cavity system applied for these measurements, we performed additional simulations, including both the longitudinal pump and higher momentum modes. In particular, we model the longitudinal probe as an external standing-wave potential with a probing frequency $\omega_\mathrm{l}$ and probing intensity $V_\mathrm{pr}$. In the frame rotating at the pump frequency, this leads to a time-dependent longitudinal potential at a frequency of $\delta_\mathrm{lt} = \omega_\mathrm{p} -\omega_\mathrm{l}$. The corresponding Heisenberg-Langevin equations are given by
\begin{align}
\frac{\partial \Psi(\mathbf{x})}{\partial t} = i\biggl[&\frac{\hbar}{2m}\nabla^2 -\frac{U_a}{\hbar}\Psi^{\dagger}_g(\mathbf{x})\Psi(\mathbf{x}) -\frac{\epsilon_{\mathrm{p}} \mathrm{cos}^2(ky)}{\hbar} -\frac{V_{\mathrm{pr}} \mathrm{cos}^2(kz)}{\hbar}-U_0\mathrm{cos}^2(kz)a^{\dagger}a\\ \nonumber
&-\sqrt{\frac{{U_0 \epsilon_{\mathrm{p}}}}{{\hbar}}}\mathrm{cos}(ky)\mathrm{cos}(kz)\left( a + a^{\dagger} \right) -\sqrt{\frac{{U_0 V_{\mathrm{pr}}}}{{\hbar}}}\mathrm{cos}^2(kz)\left( ae^{-i\delta_{\mathrm{lt}}t} + a^{\dagger}e^{i\delta_{\mathrm{lt}}t} \right)  \\ \nonumber
&-2\sqrt{\frac{\epsilon_{\mathrm{p}}V_{\mathrm{pr}}}{\hbar^2}}\mathrm{cos}(ky)\mathrm{cos}(kz)\mathrm{cos}(\delta_{\mathrm{lt}} t) \biggr]\Psi(\mathbf{x})
\end{align}
\begin{align}
&\frac{\partial a }{\partial t} = i \biggl[\Delta_{\mathrm{c}} -U_0\int \int dy\, dz\, \mathrm{cos}^2(kz) \Psi^{\dagger}_g(\mathbf{x})\Psi(\mathbf{x})\biggr]a -\kappa a + \xi \\ \nonumber
& -i \sqrt{\frac{{U_0 \epsilon_{\mathrm{p}}}}{{\hbar}}} \int \int dy\, dz\, \mathrm{cos}(ky)\mathrm{cos}(kz)  \Psi^{\dagger}_g(\mathbf{x})  \Psi(\mathbf{x})-i e^{i\delta_{\mathrm{lt}}t} \sqrt{\frac{{U_0 V_{\mathrm{pr}}}}{{\hbar}}} \int dy\, dz\, \mathrm{cos}^2(kz)  \Psi^{\dagger}_g(\mathbf{x})  \Psi(\mathbf{x}),
\end{align}
where $\Psi(\mathbf{x})$ is the atomic field operator, $a$ is the cavity mode bosonic operator, and $\xi$ is the fluctuation corresponding to the dissipation of the cavity \cite{Tuquero2024}. In the following, we use a mean-field approximation and treat both $\Psi(\mathbf{x})$ and $a$ as $c$-numbers and neglect the stochastic noise due to $\xi$. Furthermore, we use a five-point stencil to represent the kinetic energy term as done in the real-space basis simulation in Ref.~\cite{Tuquero2024}. We numerically integrate the set of coupled equations using Heun's predictor-corrector method. These simulations were performed for all the detunings shown in this manuscript, and are presented in SFigs.~\ref{fig:7} and \ref{fig:8}. Here, we obtain the maximum photon number during the dynamics of the system for various combinations of $\delta_\mathrm{lt}$, $\epsilon_\mathrm{p}$, and $\delta_\mathrm{eff}$. 
The colour maps in the left panel of SFigs.~\ref{fig:7} and \ref{fig:8} are rescaled to the maximum value of photon number obtained for the range of $\delta_\mathrm{lt}$ and $\epsilon_\mathrm{p}$ considered here. 
The excitation frequency $\omega_\mathrm{exc}$, as depicted in the right panel, for a given $\epsilon_\mathrm{p}$ and $\delta_\mathrm{eff}$ is then determined as the $\delta_\mathrm{lt}$ with the maximum photon number recorded during the dynamics. We present in the right panel of SFigs.~\ref{fig:7} and \ref{fig:8}  the simulated excitation frequency as a function of $\epsilon_\mathrm{p}$ for a selection of $\delta_\mathrm{eff}$. The dashed curve corresponds to the instantaneous LIRI prediciton given by Eq.~(1) in the main text.

\bibliography{references}